\documentclass[12pt]{article}
\usepackage{geometry}
\usepackage{times}
\usepackage[utf8]{inputenc}
\usepackage{amssymb}
\usepackage{float}

\usepackage{amsmath,amssymb,amsfonts}
\usepackage{graphicx}
\usepackage{array}
\usepackage{booktabs}
\usepackage{multicol}
\usepackage{multirow}
\usepackage{bm}
\usepackage{siunitx}
\usepackage{comment}
\usepackage{makecell}
\newcommand{\gray}[1]{\textcolor[gray]{0.5}{#1}}
\usepackage[backend=biber,style=ieee,
minbibnames=6,
maxbibnames=6,
doi=false,isbn=false,url=false,eprint=false
]{biblatex}
\defbibheading{bibliography}[\refname]{}

\DeclareSourcemap{
	\maps[datatype=bibtex, overwrite=true]{
		\map{
		    \step[fieldsource=booktitle,
			match=\regexp{.*EUSIPCO.*},
			replace={Proc. EUSIPCO}]
			\step[fieldsource=booktitle,
			match=\regexp{.*CVPR.*},
			replace={Proc. CVPR}]
			\step[fieldsource=booktitle,
			match=\regexp{.*Interspeech.*},
			replace={Proc. Interspeech}]
			\step[fieldsource=booktitle,
			match=\regexp{.*ICASSP.*},
			replace={Proc. ICASSP}]
			\step[fieldsource=booktitle,
			match=\regexp{.*ICLR.*},
			replace={Proc. ICLR}]
			\step[fieldsource=booktitle,
			match=\regexp{.*ICCV.*},
			replace={Proc. ICCV}]
			\step[fieldsource=booktitle,
			match=\regexp{.*ICML.*},
			replace={Proc. ICML}]
			\step[fieldsource=booktitle,
			match=\regexp{.*ASRU.*},
			replace={Proc. ASRU}]
			\step[fieldsource=booktitle,
			match=\regexp{.*SLT.*},
			replace={Proc. SLT}]
			\step[fieldsource=booktitle,
			match=\regexp{.*SSW.*},
			replace={Proc. SSW}]
			\step[fieldsource=booktitle,
			match=\regexp{.*WASPAA.*},
			replace={Proc. WASPAA}]
			\step[fieldsource=booktitle,
			match=\regexp{.*IJCNN.*},
			replace={Proc. IJCNN}]
            \step[fieldsource=booktitle,
            match=\regexp{.*Detection.*and.*Classification.*of.*Acoustic.*Scenes.*and.*Events.*Workshop.*},
			replace={Proc. DCASE}]
            \step[fieldsource=booktitle,
			match=\regexp{.*MLSP.*},
			replace={Proc. MLSP}]
            \step[fieldsource=booktitle,
			match=\regexp{.*ECCV.*},
			replace={Proc. ECCV}]
            \step[fieldsource=journal,
			match=\regexp{.*NeurIPS.*},
			replace={Advances in NuerIPS}]
            \step[fieldsource=journal,
			match=\regexp{.*TASLP.*},
			replace={IEEE/ACM TASLP}]
            \step[fieldsource=journal,
			match=\regexp{.*J-STSP.*},
			replace={IEEE J-STSP}]                
			\step[fieldsource=series,
			match=\regexp{.+},
			replace={{}}]
			\step[fieldsource=editor,
			match=\regexp{.+},
			replace={{}}]
			\step[fieldsource=publisher,
			match=\regexp{.+},
			replace={{}}]
			\step[fieldsource=month,
			match=\regexp{.+},
			replace={{}}]
			\step[fieldsource=location,
			match=\regexp{.+},
			replace={{}}]
			\step[fieldsource=address,
			match=\regexp{.+},
			replace={{}}]
			\step[fieldsource=organization,
			match=\regexp{.+},
			replace={{}}]
		}
	}
}

\title{Analysis and Extension of Noisy-target Training for Unsupervised Target Signal Enhancement}

\author{Takuya Fujimura \\ Graduate School of Informatics, Nagoya University, Nagoya, Japan \and Tomoki Toda \\ Information Technology Center, Nagoya University, Nagoya, Japan}

\begin{document}
\maketitle
\begin{abstract}
Deep neural network-based target signal enhancement~(TSE) is usually trained in a supervised manner using clean target signals.
However, collecting clean target signals is costly and such signals are not always available.
Thus, it is desirable to develop an unsupervised method that does not rely on clean target signals.
Among various studies on unsupervised TSE methods, Noisy-target Training~(NyTT) has been established as a fundamental method.
NyTT simply replaces clean target signals with noisy ones in the typical supervised training, and it has been experimentally shown to achieve TSE.
Despite its effectiveness and simplicity, its mechanism and detailed behavior are still unclear.
In this paper, to advance NyTT and, thus, unsupervised methods as a whole, we analyze NyTT from various perspectives.
We experimentally demonstrate the mechanism of NyTT, the desirable conditions, and the effectiveness of utilizing noisy signals in situations where a small number of clean target signals are available.
Furthermore, we propose an improved version of NyTT based on its properties and explore its capabilities in the dereverberation and declipping tasks, beyond the denoising task.
\end{abstract}

\section{Introduction}
Target signal enhancement~(TSE) is a technique to extract a target signal from a noisy observation.
In various speech communication systems, such as online meetings, hearing aids, and automatic speech recognition~(ASR) systems, this technique has been employed to extract human speech~\cite{Narayanan_2013,Yoshioka_2015,Kinoshita_2020,Fedorov_2020}.
%It plays an important role in a variety of applications such as online meetings, hearing aids, and automatic speech recognition systems~\cite{Narayanan_2013,Yoshioka_2015,Kinoshita_2020,Fedorov_2020}.
It has also been applied to various types of target signal beyond speech, including music~\cite{hennequin2020spleeter,defossez2019music,rouard2022hybrid} and environmental sounds~\cite{kavalerov2019universal,turpault2020improving,fujimura2023multi}.
This TSE technique can be classified into multi-channel and single-channel methods.
Multi-channel methods extract the target signal by leveraging the spatial information obtained from multiple microphones~\cite{DeMuth_1977,Veen_1988,Trees_2004}.
However, the physical size of the microphone array can sometimes limit its application.
Consequently, single-channel methods, which use a single microphone and perform TSE based on differences in acoustic features between the target signal and other noise signals, also play a crucial role in TSE applications.
Classical signal processing-based single-channel methods~\cite{Boll_1979,Wiener_1949} have been widely adopted owing to their simplicity and low computational cost; however, their enhancement performance is often insufficient.
In contrast, recent single-channel TSE methods have achieved significant performance improvements by incorporating deep neural networks~(DNNs)~\cite{Williamson_2016,luo2019conv,subakan2021attention,Koizumi_2021,wang2023tf,rouard2022hybrid}.

\begin{figure*}[t!]
  \centering
  \includegraphics[width=0.99\linewidth]{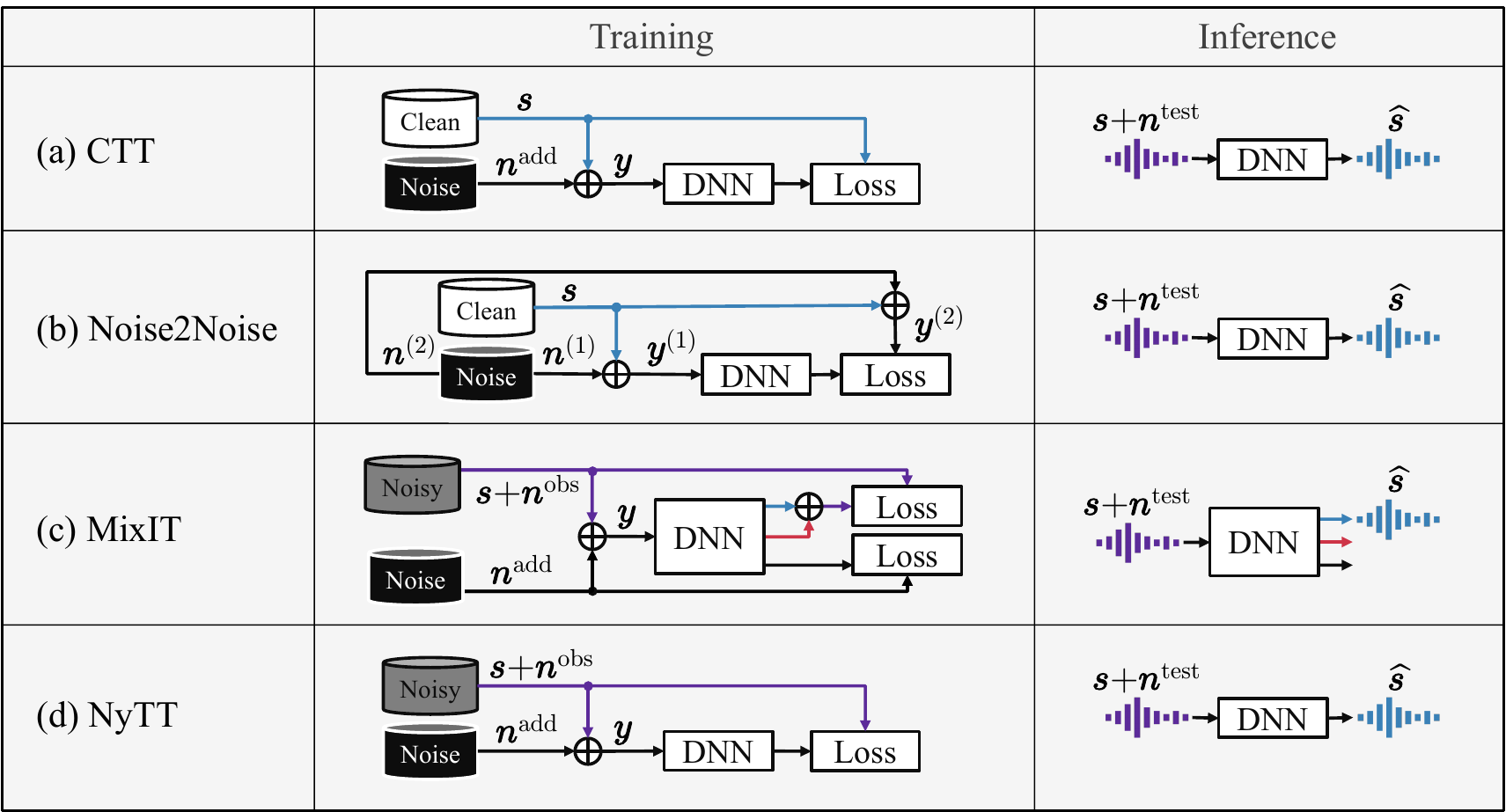}
  \caption{Comparison of NyTT and its related methods.}
\label{fig:comparison}
\end{figure*}

Most single-channel DNN-based TSE methods rely on supervised learning with clean target signals, which we refer to as Clean-target Training (CTT) (Fig.~\ref{fig:comparison}(a)).
%\footnote{In the field of general machine learning, parallel training is identical to supervised training. However, in the field of TSE, parallel training without clean target signal is also treated as unsupervised training.~\cite{Wisdom_2020}. In this paper, we use ``parallel training'' instead of ``supervised training'' to avoid ambiguity.}
In CTT, we input a noisy signal into a DNN and train it to predict the corresponding clean target signal.
CTT is an appropriate strategy, and various improvements have been made, including modifications to model architectures (e.g., convolutional networks~\cite{luo2019conv}, recurrent networks~\cite{luo2020dual}, and Transformers~\cite{subakan2021attention,Koizumi_2021}), loss functions (e.g., mean-squared-error (MSE) and signal-to-noise ratio (SNR)~\cite{erdogan18_interspeech}), and signal representations to which TSE is applied (e.g., amplitude spectrograms~\cite{lu2013speech}, complex spectrograms~\cite{Williamson_2016,wang2023tf}, waveforms~\cite{pascual2017segan,luo2019conv,subakan2021attention}, and both spectrograms and waveforms~\cite{rouard2022hybrid}).
Furthermore, memory-efficient~\cite{tzinis2020sudo,subakan2022resource} and real-time~\cite{defossez2020real} architectures have also been explored.
Despite these improvements, CTT still has one major problem: collecting clean target signals is costly.
Typically, clean target signals are recorded in controlled settings, such as an anechoic chamber, to prevent degradation from environmental noise and reverberation.
Consequently, the recording process is costly and time-consuming, limiting the amount of training data. %, which limits its practical applications. 
%While it is theoretically possible to achieve TSE of any target signals by altering the training target signal data, recording clean target signals for environmental sounds, such as those from animals and vehicles, is often not feasible.
Moreover, although it is theoretically possible to achieve the TSE of any target signals, such as animals and vehicle sounds, it is often not feasible to record such clean target signals.
Therefore, the types of target signal used in CTT are realistically limited.
% However, recording clean target signals of environmental sounds such as animal and vehicle sounds is not feasible.
% Therefore, the type of the target signal in CTT has been also realistically limited.

To alleviate this limitation, unsupervised\footnote{In the context of TSE, methods that do not require clean target signals are considered unsupervised, even if the training is performed in a supervised manner~\cite{Wisdom_2020}.} TSE methods have been studied~\cite{Wisdom_2020,ito2023audio,Fu_2022,kashyap2021speech,alamdari2021improving,Fujimura_2021}.
PULSE~\cite{ito2023audio} is an unsupervised TSE method based on positive-unlabeled~(PU) learning~\cite{du2014analysis} and uses noisy target signals and additional noise signals for its training.
PU learning is a machine learning technique that enables the classification of positive and negative examples using positive and unlabeled training data.
On the basis of this technique, PULSE classifies local patches of amplitude spectrograms into noise (positive) or target signal (negative) components.
For training, patches of noise signals are used as positive data, while patches of noisy signals are used as unlabeled data since a noisy signal contains both noise and target signal patches.
%Using patches of noise signals as positive data and patches of noisy signals as unlabeled data, PULSE trains a DNN to classify local patches of amplitude spectrograms into noise (positive) or target signal (negative) components.
%, where patches of noisy signals are treated as unlabeled data, as they contain both noise and target signal patches.
During inference, PULSE performs TSE by applying a mask to the input amplitude spectrogram, filtering out noise (positive) patches.
Another method, MetricGAN-U, is based on a generative adversarial network (GAN) and uses only noisy signals for its training.
In MetricGAN-U, the discriminator is trained to predict a signal quality metric, whereas the generator is trained to maximize the evaluation from the discriminator.
This achieves the unsupervised TSE by employing a non-intrusive metric, which does not use a clean target signal as a reference, as the metric that the discriminator mimics.
Although PULSE and MetricGAN-U have demonstrated their TSE capabilities, they cannot directly inherit advancements made in the CTT framework owing to their specialized training algorithms.
%limitations due to their specialized training algorithms, which prevent them from directly inheriting advancements made in the CTT framework.
For example, PULSE restricts TSE models to the time--frequency (T--F) masking approach because it relies on the classification of spectrogram patches, even though time-domain models have also been developed within the CTT framework~\cite{luo2019conv,subakan2021attention,Koizumi_2021,rouard2022hybrid}.
Moreover, MetricGAN-U requires a non-intrusive evaluation metric for its training, but it is not always available.
Although a pre-trained DNN-based evaluation metric predictor can be employed as the non-intrusive metric, as demonstrated in the experiments in \cite{Fu_2022}, constructing this predictor still requires clean target signals.
Thus, it does not serve as an essential solution, especially when developing TSE systems for new types of target signal.

In contrast to the aforementioned unsupervised TSE methods, another approach utilizes noisy signals in the same training algorithms as CTT~\cite{Wisdom_2020,kashyap2021speech,Fujimura_2021}.
Noisy-target Training (NyTT)~\cite{Fujimura_2021} is the basic method in this approach (Fig.~\ref{fig:comparison}(d)).
NyTT utilizes a noisy signal as the target signal instead of a clean one.
It trains a DNN to predict the noisy target from a signal synthesized by mixing the noisy target with additional noise.
During inference, the enhanced signal can be obtained directly from the DNN by inputting an unprocessed noisy signal.
NyTT has been experimentally shown to achieve TSE without clean target signals and has the same training algorithm as CTT, which allows us to easily inherit advancements made in the CTT framework.
However, the exact mechanism and desirable conditions of NyTT have not been clarified, hindering further advancements.

In this paper, we aim to advance the field of unsupervised TSE by analyzing the fundamental method, NyTT, from various perspectives and deepening our understanding of NyTT\footnote{This paper is an extension of our previous paper~\cite{Fujimura_2023}. Compared with the previous paper, this paper provides a more thorough discussion of related works, a more comprehensive analysis of the desirable conditions for NyTT, and an investigation of its capability in dereverberation and declipping tasks, employing multiple evaluation metrics.}.
Through this analysis, we clarify 1) the mechanism of NyTT,
2) the desirable conditions of NyTT, and
3) the effectiveness of utilizing noisy signals in situations where a small number of clean target signals are available.
Additionally, 4) we propose an improved version of NyTT based on its properties, demonstrating its potential to achieve performance comparable to CTT by iteratively improving the quality of the noisy target signals.
Finally, 5) we demonstrate that NyTT can also handle dereverberation and declipping tasks, inheriting the broad applicability of CTT.

The rest of this paper is organized as follows:
Sec.~\ref{sec:related} provides details of NyTT and its closely related works.
In Sec.~\ref{sec:motivation}, we outline the motivation and contents of our analyses.
In Secs.~\ref{sec:denoise},~\ref{sec:dereverb}, and \ref{sec:declip}, we present experimental results in denoising, dereverberation, and declipping tasks, respectively.
Finally, in Sec.~\ref{sec:conclusion}, we conclude this paper.

\section{NyTT and its related works}
\label{sec:related}
%In this section, we explain NyTT and its closely related works.

\subsection{Noise2Noise}
Noise2Noise~\cite{Lehtinen_2018} is an unsupervised training method originally proposed for image denoising (Fig.~\ref{fig:comparison}(b)).
In Noise2Noise, pairs of noisy signals $(\bm{y}^{(1)}=\bm{s}+\bm{n}^{(1)}, \bm{y}^{(2)}=\bm{s}+\bm{n}^{(2)})$ are used as training data, where the two noisy signals $\bm{y}^{(1)}$ and $\bm{y}^{(2)}$ share the same clean target signal $\bm{s}$ but have different noise components $\bm{n}^{(1)}$ and $\bm{n}^{(2)}$.
A DNN $f(\cdot)$ is trained to minimize the following prediction error:
\begin{equation}
  \label{eq:n2n1}
    \mathcal{L}^{\rm N2N} = \mathbb{E}_{(\bm{y}^{(1)}, \bm{y}^{(2)})\sim\mathcal{D}}[L(f(\bm{y}^{(1)};\theta), \bm{y}^{(2)})],
    % ||f(\bm{y}_1)-\bm{y}_2||^2_2.
\end{equation}
where $\mathbb{E}[\cdot]$ is the expectation operator, $\mathcal{D}$ is a training dataset, $L(\cdot)$ is a loss function, and $\theta$ is the set of parameters of the DNN $f(\cdot)$.
Here, Noise2Noise trains the DNN to acquire one-to-one mappings between the two noisy signals $\bm{y}^{(1)}$ and $\bm{y}^{(2)}$.
However, multiple plausible outputs can exist for a given input, especially when there is no consistent relationship between the two noise signals $\bm{n}^{(1)}$ and $\bm{n}^{(2)}$.
In such a case, if the loss function is MSE, the optimal solution becomes the average of the plausible candidates.
%; therefore, zero-mean noise can be removed from the output signals even when using noisy signals as the target signals.
For example, we consider the optimal output $\hat{\bm{y}}^{(2)}$ for a given input $\bm{y}^{(1)}$, when using MSE as the loss function.
Here, the training objective is to minimize the following $\mathcal{L}^{\rm N2N}_{\bm{y}^{(2)}|\bm{y}^{(1)}}$:
\begin{align}
    \mathcal{L}^{\rm N2N}_{\bm{y}^{(2)}|\bm{y}^{(1)}} &= \mathbb{E}_{\bm{y}^{(2)}|\bm{y}^{(1)}}[L(\hat{\bm{y}}^{(2)}, \bm{y}^{(2)})].
     %&= \mathbb{E}_{\bm{y}^{(2)}|\bm{y}^{(1)}}\left [\|\hat{\bm{y}}^{(2)} - \bm{y}^{(2)} \|_2^2 \right ],\\
    % ||f(\bm{y}_1)-\bm{y}_2||^2_2.
\end{align}
Therefore, $\hat{\bm{y}}^{(2)}$ is obtained as
\begin{align}
  \frac{\partial}{\partial \hat{\bm{y}}^{(2)}}\mathcal{L}^{\rm N2N}_{\bm{y}^{(2)}|\bm{y}^{(1)}}&=0,\\
  \frac{\partial}{\partial \hat{\bm{y}}^{(2)}}\mathbb{E}_{\bm{y}^{(2)}|\bm{y}^{(1)}}\left [\|\hat{\bm{y}}^{(2)} - \bm{y}^{(2)} \|_2^2 \right ]&=0,\\
  % 2\mathbb{E}_{\bm{y}^{(2)}|\bm{y}^{(1)}}\left [\hat{\bm{y}}^{(2)} - \bm{y}^{(2)} \right ]&=0,\\
  \hat{\bm{y}}^{(2)} &= \mathbb{E}_{\bm{y}^{(2)}|\bm{y}^{(1)}} [ \bm{y}^{(2)} ].
\end{align}
%where we can see that $\hat{\bm{y}}^{(2)}$ becomes the expectation of the clean target signal $\bm{s}$.
%This result shows that $\hat{\bm{y}}^{(2)}$ is the average of the plausible candidates.
%, and the DNN $f(\cdot)$ is also trained to output it.
% the DNN $f(\cdot)$ is trained to output the average of the plausible candidates if the loss function is the mean-squared-error (MSE).
This averaging effect is observed as a problem of a blurred output in super-resolution and a greyish output in autocoloring~\cite{Ledig_2017,Zhang_2016,Isola_2017}.
On the basis of this property, Noise2Noise can achieve the same denoising training as CTT without requiring clean target signals since the averaging effect can remove zero-mean noise in the output signal (i.e., $\hat{\bm{y}}^{(2)} = \mathbb{E}_{\bm{y}^{(2)}|\bm{y}^{(1)}} [ \bm{s} + \bm{n}^{(2)} ]=\mathbb{E}_{\bm{y}^{(2)}|\bm{y}^{(1)}} [ \bm{s}]$).

It has been theoretically and experimentally proven that Noise2Noise can achieve denoising training using only noisy signals.
In the case of images, it is relatively easy to obtain pairs of noisy signals, $(\bm{y}^{(1)}, \bm{y}^{(2)})$, that share the same clean image $\bm{s}$ by taking consecutive shots when the subject is static.
In this case, Noise2Noise is a useful technique.
However, in the case of audio, it is not possible to naturally obtain such pairs of noisy signals.
Instead, we must synthesize them using a clean audio signal $\bm{s}$.
Therefore, in the audio TSE task, Noise2Noise does not serve as an essential solution~\cite{kashyap2021speech,alamdari2021improving}.

% \begin{align}
%   \label{eq:n2n2}
%   \begin{split}
%     \mathcal{L}^{\rm N2N} &= \mathbb{E}\left[||f(\bm{y}^{(1)}; \theta)-\bm{y}^{(2)}||^2_2\right]\\
%     &= \mathbb{E}\left[||f(\bm{y}^{(1)}; \theta)-\bm{s}-\bm{n}^{(2)}||^2_2\right]\\
%     &= \mathbb{E}\left[||f(\bm{y}^{(1)}; \theta)-\bm{s}||^2_2\right]
%     \\&-2\mathbb{E}\left[(f(\bm{y}^{(1)}; \theta)-\bm{s})\bm{n}^{(2)}\right]
%     +\mathbb{E}\left[||\bm{n}^{(2)}||^2_2\right],
%   \end{split}
% \end{align}
% where we omitted $\mathbb{E}_{(\bm{y}^{(1)}, \bm{y}^{(2)})\sim\mathcal{D}}[\cdot]$ to $\mathbb{E}[\cdot]$ for simplicity.
% Here, when the noise signals $\bm{n}^{(2)}$ have zero-mean distribution and it is uncorrelated with $f(\bm{y}^{(1)}; \theta)-\bm{s}$, the second term becomes zero.

\subsection{MixIT}
MixIT~\cite{Wisdom_2020} is an unsupervised training method for sound source separation, and it can also be used for TSE (Fig.~\ref{fig:comparison}(c)).
In MixIT, training is conducted using noisy signals $\bm{x}=\bm{s}+\bm{n}^{\rm obs}$ and additional noise signals $\bm{n}^{\rm add}$, where $\bm{n}^{\rm obs}$ represents the noise already present in $\bm{x}$ at the time of observation.
MixIT trains a DNN to minimize the following prediction error $\mathcal{L}^{\rm MixIT}$:
\begin{align}
  \label{eq:mixit}
      \mathcal{L}^{\rm MixIT} &= \mathbb{E}_{(\bm{x},\,\bm{n}^{\rm{add}})\sim\mathcal{D}}\left[\min (\mathcal{L}_{\rm MixIT1}, \mathcal{L}_{\rm MixIT2})\right],\\
      \mathcal{L}_{\rm MixIT1} &= L( \bm{u}_1+\bm{u}_2, \bm{x})+L( \bm{u}_3, \bm{n}^{\rm add}),\\
      \mathcal{L}_{\rm MixIT2} &= L( \bm{u}_1+\bm{u}_3, \bm{x})+L( \bm{u}_2, \bm{n}^{\rm add}).
\end{align}
Here, $\bm{u}_1$, $\bm{u}_2$, and $\bm{u}_3$ are the outputs of the DNN $f(\bm{y};\theta)$, where $\bm{y}=\bm{x}+\bm{n}^{\rm add}$.
MixIT can achieve TSE because $(\bm{u}_1, \bm{u}_2, \bm{u}_3)=(\bm{s}, \bm{n}^{\rm obs}, \bm{n}^{\rm add})$ or $(\bm{s}, \bm{n}^{\rm add}, \bm{n}^{\rm obs})$ is the optimal solution for $\mathcal{L}^{\rm MixIT}$.
Although $\mathcal{L}^{\rm MixIT}$ can also be minimized by outputting a noisy signal as $\bm{u}_1$ (i.e., $(\bm{u}_1, \bm{u}_2, \bm{u}_3) = (\bm{x}, \bm{0}, \bm{n}^{\rm add})$), this issue can be avoided under the assumption that $\bm{y}$ does not provide any information indicating that $\bm{x}$ consists of $\bm{s}$ and $\bm{n}^{\rm obs}$.
Under this assumption, the DNN cannot estimate a pair of signals that compose $\bm{x}$ and, therefore, cannot always accurately estimate $\bm{x}$.
Consequently, the DNN is trained to separate $\bm{y}$ into individual sources, as this always minimizes $\mathcal{L}^{\rm MixIT}$.
When this assumption does not hold (e.g., when the characteristics of $\bm{n}^{\rm obs}$ and $\bm{n}^{\rm add}$ differ and are distinguishable), it has been observed that MixIT suffers from performance degradation~\cite{saito2021training,maciejewski2021training}.
% In practice, MixIT suffers from performance degradation when it separates only $\bm{n}^{\rm add}$ from $\bm{y}$, especially when the characteristics of $\bm{n}^{\rm obs}$ and $\bm{n}^{\rm add}$ differ~\cite{saito2021training,maciejewski2021training}.
%lead to cases where both $\bm{s}+\bm{n}^{\rm obs}$ (which produces a low loss) and $\bm{s}+\bm{n}^{\rm add}$ (which produces a high loss) are output as $\bm{u}_1$.

MixIT is one of the major unsupervised TSE methods, and several improvements have been proposed.
For instance, one method mitigated the overseparation problem by introducing a penalty term for the number of active sources and the correlation between the output sources~\cite{wisdom2021sparse}.
Other methods produced better separation by using a pre-trained classification model (e.g., an audio event classification or an ASR model)~\cite{wisdom2021sparse,trinh2022unsupervised} or by employing a loss function that relaxes the training difficulty~\cite{saito2021training,maciejewski2021training}.
Furthermore, the teacher--student learning approach has also been adopted~\cite{zhang2021teacher,Tzinis_2022,karamatli2022mixcycle,saijo2023self,li2024remixed2remixed}.
In this approach, the student model is trained using the outputs of the teacher model pre-trained by MixIT as the pseudo-target signals.

% For example, some works have tackled the over-separation problem, which is the problem of over-separating a target signal due to the training a DNN to separate mixture as much as possible~\cite{Wisdom_2020,wisdom2021sparse,zhang2021teacher,karamatli2022mixcycle}.
% In \cite{wisdom2021sparse}, over-separation problem has been mitigated by introducing penalty term for the number of active sources and correlation between the output sources.
% In addition, MixIT has been improved by jointly using a pre-trained classification model (am audio event classification or an ASR model)~\cite{wisdom2021sparse,trinh2022unsupervised}, and introducing a loss function that relaxes the training difficulty~\cite{saito2021training,maciejewski2021training}.
% Furthermore, MixIT has been combined with a Teacher-Student learning to improve the performance for over-separation problem or domain adaptation~\cite{zhang2021teacher,Tzinis_2022,karamatli2022mixcycle,saijo2023self,li2024remixed2remixed}.

\subsection{NyTT}

NyTT~\cite{Fujimura_2021} is an unsupervised training method designed for TSE (Fig.~\ref{fig:comparison}(d)).
In NyTT, noisy signals $\bm{x}$ and additional noise signals $\bm{n}^{\rm add}$ are used as training data, and a \textit{more noisy} signal $\bm{y}$ is generated as $\bm{y}=\bm{x}+\bm{n}^{\rm add}$.
NyTT trains a DNN to minimize the following prediction error $\mathcal{L}^{\rm NyTT}$:
\begin{align}
  \label{eq:nytt}
      \mathcal{L}^{\rm NyTT} = \mathbb{E}_{(\bm{y},\,\bm{x})\sim\mathcal{D}}\left[L( f(\bm{y}; \theta), \bm{x})\right].
\end{align}
NyTT was inspired by Noise2Noise and realizes Noise2Noise training in the TSE task by considering $\bm{y} = \bm{s} + (\bm{n}^{\rm obs} + \bm{n}^{\rm add}) = \bm{s} + \bm{n}^{(1)}$ and $\bm{x} = \bm{s} + \bm{n}^{\rm obs} = \bm{s} + \bm{n}^{(2)}$ as pairs of noisy signals.
The prediction error is calculated using MSE in the time domain, under the assumption that the noise has a zero-mean distribution.
Despite the lack of theoretical proof, NyTT has been experimentally demonstrated to achieve TSE without clean target signals~\cite{Fujimura_2021}.

Although NyTT and MixIT stem from different conceptual foundations, their resulting training algorithms are similar.
The primary difference is which output is selected as the enhanced signal during the inference.
MixIT involves a separation task, where $\bm{y}$ is separated into several sound sources, and one of them is selected as the enhanced signal.
In contrast, NyTT has only one slot and uses the output as the enhanced signal.
Therefore, NyTT can be viewed as the simplified version of MixIT, where the DNN is trained with only $L(\bm{u}_1+\bm{u}_2, \bm{x})$ and the enhanced signal is created by $\bm{u}_1+\bm{u}_2$ during the inference.
Both MixIT and NyTT are major unsupervised TSE methods, and they provide a greater flexibility than specialized unsupervised training algorithms.
However, as mentioned above, NyTT has a simpler and more flexible architecture than MixIT.
The simple architecture enables us to make improvements and expansions easily, and it is more suitable for analysis.
For these reasons, we have chosen NyTT as the target of our analysis.

One limitation of NyTT is that it requires the zero-mean noise assumption and the use of MSE as conditions for Noise2Noise.
Conversely, if NyTT does not require these conditions, it can be easily applied to wider ranges of tasks and loss functions.
In Sec.~\ref{sec:validity}, we demonstrate that NyTT indeed works without these conditions, and in Secs.~\ref{sec:dereverb} and \ref{sec:declip}, we further show its effectiveness in dereverberation and declipping tasks.
%In contrast, if NyTT works without these conditions, it can be easily applied to a wider range of tasks and loss functions.
%Therefore, in order to extend its potential, we demonstrate that NyTT works without these conditions in Sec.~\ref{sec:validity}, it also works in dereverberation and declipping tasks in Sec.~\ref{sec:dereverb} and \ref{sec:declip}.

% NyTT was proposed inspired by Noise2Noise, which considers the $\bm{y}=\bm{s}+(\bm{n}^{\rm obs}+\bm{n}^{\rm add})=\bm{s}+\bm{n}^{(1)}$ and $\bm{x}=\bm{s}+\bm{n}^{\rm obs}= \bm{s}+\bm{n}^{(2)}$ as the pairs of noisy signals.

\section{Motivation and content of the investigation}
\label{sec:motivation}
\subsection{Validity of the interpretation of NyTT}
NyTT has been proposed, inspired by Noise2Noise, which utilizes noisy signals as target signals on the basis of the averaging effect of MSE loss function.
On the other hand, NyTT can also be interpreted as being trained to remove $\bm{n}^{\rm add}$ from the \textit{more noisy} signal $\bm{y}$, performing TSE by removing noise components corresponding to $\bm{n}^{\rm add}$ from the noisy input signal.
Therefore, we investigate whether NyTT can be interpreted as Noise2Noise or not, through 1) the analysis of the signals processed in NyTT (Sec.~\ref{sec:validity1}) and 2) the evaluation of NyTT with a loss function that does not satisfy the conditions of Noise2Noise (Sec.~\ref{sec:validity2}).
If NyTT is not Noise2Noise, the zero-mean noise assumption and the use of MSE will no longer be necessary, allowing us to use various loss functions and apply NyTT to various tasks.

\begin{figure}[t!]
  \centering
  \includegraphics[width=0.99\columnwidth]{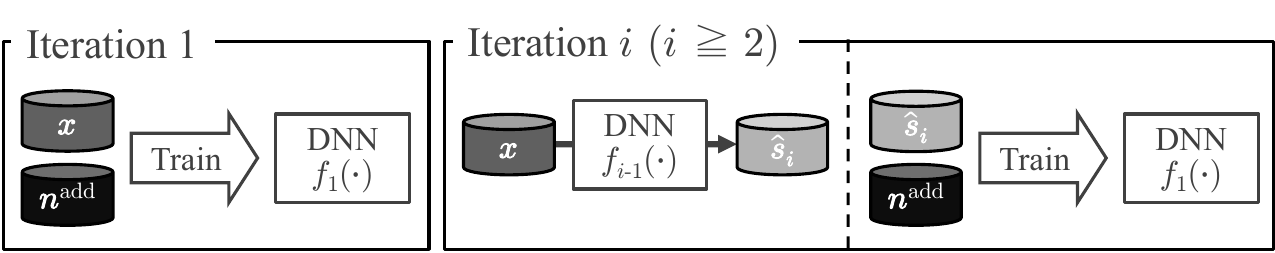} 
  \caption{Overview of IterNyTT.}
\label{fig:iternytt_overview}
\end{figure}

\subsection{Improvement of NyTT through iteration}
It has been shown that the performance of NyTT improves as the SNR of the noisy target increases~\cite{Fujimura_2021}.
On the basis of this property, we propose IterNyTT, which achieves better performance through an iterative process (Fig.~\ref{fig:iternytt_overview}).
In the first iteration of IterNyTT, we train a DNN $f_1(\cdot)$ using NyTT with the noisy target $\bm{x}$.
Next, we apply TSE to the noisy target $\bm{x}$ using $f_1(\cdot)$ and obtain the enhanced signal $\hat{\bm{s}}_{i-1}$.
Then, we train another DNN $f_2(\cdot)$ using NyTT with $\hat{\bm{s}}_{i-1}$ as the noisy target signal.
To prevent the degradation of the target signal component, we apply TSE to the original noisy target $\bm{x}$, not to the already enhanced signal $\hat{\bm{s}}_{i-1}$.
Through this iterative process, we can improve the SNR of noisy targets and expect the performance improvement of NyTT.
We investigate the effectiveness of IterNyTT in Sec.~\ref{sec:iternytt}.

\subsection{Effects of mismatches between noise signals}
In the NyTT framework, there are three types of noise signal: $\bm{n}^{\rm obs}$, $\bm{n}^{\rm add}$, and noise included in the test data $\bm{n}^{\rm test}$.
In Sec.~\ref{sec:noise}, we investigate the effects of mismatches between types of noise signal on the performance of NyTT and IterNyTT.
For instance, the performance of CTT is degraded when there is a mismatch between $\bm{n}^{\rm add}$ and $\bm{n}^{\rm test}$~\cite{Wisdom_2020,ito2023audio}.
%Similarly, we expect that the mismatch between three types of noise in NyTT affects the performance.
In the experiments, 1) we evaluate the performance of CTT, NyTT, and IterNyTT under mismatched conditions, and investigate the effects of each mismatch (Sec.~\ref{sec:noise1}).
Additionally, considering the effects of mismatch between $\bm{n}^{\rm obs}$ and $\bm{n}^{\rm test}$, we investigate the effects of 2) the SNRs of the noisy targets $\bm{x}$ ($\mathrm{SNR}_{\bm{x}}=\log_{10}||\bm{s}||^2_2/||\bm{n}^{\rm obs}||^2_2$) and 3) the SNRs of the \textit{more noisy} signals $\bm{y}$ ($\mathrm{SNR}_{\bm{y}}=\log_{10}||\bm{x}||^2_2/||\bm{n}^{\rm add}||^2_2$) on performance.

\subsection{Effectiveness of utilizing noisy signals in a situation where clean target signals are available}
Collecting a large number of clean target signals is challenging owing to high recording costs.
In some cases, no clean target signals may be available, whereas in others, a small number of clean target signals are available, depending on the task.
%Depending on the task, there may be situations where no clean target signals are available, while in other cases, a small amount of clean target signals can be obtained.
%In fact, for human speech, clean speech corpora are available~\cite{libri,timit}.
In Sec.~\ref{sec:volume}, assuming situations where a small number of clean target signals are available, we investigate the effectiveness of utilizing a larger number of noisy signals.
% as noisy signals can be easily collected.
%We conduct evaluations under conditions where a small amount of clean target signals and a larger amount of noisy signals are available, as noisy signals can be easily collected.

\subsection{Capabilities in the dereverberation task}
The dereverberation task aims to restore an original signal from a reverberant signal.
CTT can achieve dereverberation in the same manner as the denoising task by inputting reverberant signals and training a DNN to predict clean target signals.
In Sec.~\ref{sec:dereverb}, we investigate whether NyTT can also perform dereverberation as CTT does.
Specifically, we evaluate the performance of CTT, NyTT, and IterNyTT, where CTT predicts a clean target signal $\bm{s}$ from a reverberant signal $\bm{s} \ast \bm{r}^{\rm add}$ and NyTT predicts $\bm{x}= \bm{s} \ast \bm{r}^{\rm obs}$ from $\bm{x} \ast \bm{r}^{\rm add}$, where $\bm{r}^{\rm obs}$ and $\bm{r}^{\rm add}$ are room impulse responses (RIRs), and $\ast$ denotes the convolution operation.

\subsection{Capabilities in the declipping task}
In addition to the dereverberation task, we investigate the capabilities of NyTT in the declipping task.
The declipping task aims to restore an original signal from a clipped signal where the clipping function $f_\mathrm{clip}:\mathbb{R}^T\rightarrow\mathbb{R}^T$ is defined as
\begin{align}
  \label{eq:clip}
      f_\mathrm{clip}(\bm{s}; c)[m] = \left\{
\begin{array}{ll}
\bm{s}[m] & |\bm{s}[m]|<c \\
c\cdot\mathrm{sgn}(\bm{s}[m]) & \textrm{otherwise}
\end{array}
\right.\,\,,
\end{align}
where $m$ is the time index and $c$ is a clipping threshold.
In Sec.~\ref{sec:declip}, we evaluate the performance of CTT, NyTT, and IterNyTT, where CTT predicts a clean target signal $\bm{s}$ from a clipped signal $f_\mathrm{clip}(\bm{s};c^{\rm add})$ NyTT predicts $\bm{x}=f_\mathrm{clip}(\bm{s};c^{\rm obs})$ from $f_\mathrm{clip}(\bm{x};c^{\rm add})$, and $c^{\rm add}<c^{\rm obs}$.

% \begin{table*}[tt]
% \caption{Dataset used in the denoising task}
% \begin{center}
% \begingroup
% \renewcommand{\arraystretch}{1.4}
% \begin{tabular}{l|l|l|l|c} 
% \toprule
%     Signal type & Original Dataset & \multicolumn{2}{l|}{Train set} & Test set\\
%     \midrule
%     \makecell[{{l}}]{clean target signal $\bm{s}$ \\ (Utterances)} & LibriSpeech~\cite{libri} & \multicolumn{2}{c|}{(10,000)} &(1,000) \\
%     %\midrule
%     \hline
%     \multirow{3}{*}{\makecell[{{l}}]{Noise $\bm{n}$ \\ (Volume [hr])}} & CHiME3~\cite{chime3} & \makecell[{{l}}]{\texttt{CHiME-A} \\(3.92)} & \makecell[{{l}}]{\texttt{CHiME-B} \\(3.92)} & \makecell[{{l}}]{\texttt{CHiME-C}\\ (0.56)}\\
%     \cline{2-5}
%     & \makecell[{{l}}]{VoiceBank-\\DEMAND~\cite{vbd}} & \makecell[{{l}}]{\texttt{DEMAND-A}\\ (4.70)} & \makecell[{{l}}]{\texttt{DEMAND-B} \\(4.69)}& \\ 
%     \cline{2-5}
%     & \makecell[{{l}}]{DCASE 2016\\ Task2 ~\cite{dcase}} & \multicolumn{2}{c|}{\texttt{DCASE} (0.07)} & \\
% \bottomrule
% \end{tabular}
% \endgroup
% \end{center}
% \label{tbl:dataset}
% \end{table*}

\begin{table*}[tt]
  \caption{Datasets used in the denoising task}
  \begin{center}
  \begingroup
  \renewcommand{\arraystretch}{1.2}
  \resizebox{\columnwidth}{!}{
  \begin{tabular}{l|l|l|l|c} 
  \toprule
      Signal type & Original Dataset & \multicolumn{2}{l|}{Train set} & Test set\\
      \midrule
      \makecell[{{l}}]{Clean signal $\bm{s}$ \\ (Utterances)} & LibriSpeech~\cite{libri} & \multicolumn{2}{c|}{(10,000)} &(1,000) \\
      %\midrule
      \hline
      \multirow{3}{*}{\makecell[{{l}}]{Noise $\bm{n}$ \\ (Volume [h])}}  &CHiME3~\cite{chime3} & \texttt{CHiME-A} (3.92) & \texttt{CHiME-B} (3.92) & \texttt{CHiME-C} (0.56)\\
      \cline{2-5}
      & VoiceBank-DEMAND~\cite{vbd} & \texttt{DEMAND-A} (4.70) & \texttt{DEMAND-B} (4.69)& \\ 
      \cline{2-5}
      & DCASE 2016 Task2 ~\cite{dcase} & \multicolumn{2}{c|}{\texttt{DCASE} (0.07)} & \\
  \bottomrule
  \end{tabular}
  }
  \endgroup
  \end{center}
  \label{tbl:dataset}
  \end{table*}

\section{Experimental analysis in the denoising task}
\label{sec:denoise}
One promising application of unsupervised TSE methods is the extraction of environmental sounds, as clean human speech corpora have already been created through extensive community efforts~\cite{libri,timit}. However, in our experiments, we used clean human speech as the target signal, as this allows us to control the quality and volume of the noisy signals by distorting the clean speech corpora. We believe that the insights gained from our experiments are equally applicable to other types of target signal.

\subsection{Setups}
In the experiments, we used several datasets as shown in Table~\ref{tbl:dataset}, including LibriSpeech~\cite{libri}, CHiME3~\cite{chime3}, noise extracted from the training dataset of VoiceBank-DEMAND~\cite{vbd}, and the training dataset of the DCASE 2016 Challenge Task2 dataset~(\texttt{DCASE})~\cite{dcase}.
CHiME3 included background noise recorded in a bus, a cafe, a pedestrian area, and a street junction.
VoiceBank-DEMAND included noise recorded in a kitchen, an office, a cafe, and a subway, along with artificially synthesized bubble and white noise.
The DCASE 2016 Task2 dataset included sounds of coughing, door knocking, and telephone ringing.
Thus, there are differences in the types of noise across these three datasets.
The noise signals from CHiME3 were segmented every \SI{10}{s}, and \SI{7.83}{h} of data were split into \texttt{CHiME-A} and \texttt{CHiME-B}, and another \SI{0.56}{h} of data were used as \texttt{CHiME-C}.
11,572 clips of VoiceBank-DEMAND were split into two subsets, \texttt{DEMAND-A} and \texttt{DEMAND-B}.
The noisy target training dataset was generated by mixing 10,000 utterances of clean target signals from LibriSpeech and noise signals $\bm{n}^{\mathrm{obs}}$ at $\mathrm{SNR}_{\bm{x}}$ randomly selected from 0, 5, 10, and 15~dB.
The test dataset was generated by mixing 1,000 utterances of clean target signals from LibriSpeech and \texttt{CHiME-C} at SNR randomly selected from 2.5, 7.5, 12.5, and 17.5~dB.
During training, the input signal to the DNN was generated by mixing target signals and additional noise signals, and the $\mathrm{SNR}_{\bm{y}}$ ranges were -5 to 5~dB for NyTT, and 0, 5, 10, and 15~dB for IterNyTT after the second iteration and CTT.
We evaluated the performance using the best validation epoch, where the validation was conducted with 50 pairs of input and target signals generated under the same $\mathrm{SNR}_{\bm{y}}$, $\bm{n}^{\rm obs}$, and $\bm{n}^{\rm add}$ as in the training dataset.
%We refer to the dataset of the noisy signals as ``clean set name\texttt{-}noise set name'', i.e., \texttt{Libri1-CHiME1} means a mixture set of \texttt{Libri1} and \texttt{CHiME1}.

The DNN architecture was CNN-BLSTM used in ~\cite{Kawanaka_2020}.
The input feature was the log-amplitude spectrogram, and the network estimated a complex-valued T--F mask.
For the short-time Fourier transform~(STFT) parameters, the frame shift, window size, and DFT size were set to 128, 512, and 512 samples, respectively, using the Hamming window with a sampling frequency of 16 kHz.
We trained the DNN for 1,500 epochs with a mini-batch size of 50, using the Adam optimizer~\cite{Kingma_2015} with a fixed learning rate of 0.0001.
For the loss function, we used MSE calculated in the time domain.
As the metrics, we used the scale-invariant signal-to-distortion ratio~(SI-SDR)~\cite{Roux_2019}, the perceptual evaluation of speech quality~(PESQ)~\cite{pesq}, and the short-time objective intelligibility~(STOI)~\cite{taal2011algorithm}.

\subsection{Validity of interpretation of NyTT}
\label{sec:validity}
In this section, we trained the DNN using \texttt{CHiME-A} and \texttt{CHiME-B} as $\bm{n}^{\rm obs}$ and $\bm{n}^{\rm add}$, respectively.
%, and analyzed its behavior.

%In this part, to confirm whether NyTT is Noise2Noise or not, we conducted two analyses:
%(1) an analysis of signals handled in NyTT and 
%(2) an evaluation of NyTT with a loss function that does not meet the conditions of Noise2Noise.

\subsubsection{Analysis of signals processed in NyTT}
\label{sec:validity1}
If NyTT is Noise2Noise, the output signal corresponding to the \textit{more noisy} signal $\bm{y}$ should be the estimate of the clean target signal $\bm{s}$.
To investigate this, we analyzed the output signals when we input $\bm{y}$ to the DNN.
To analyze the output signals during training, we generated \textit{more noisy} signals by using the training dataset.
Additionally, to analyze the output signals for unseen \textit{more noisy} signals, we generated \textit{more noisy} signals by mixing noisy signals from the test dataset and \texttt{CHiME-C} at SNRs ranging from -5 to 5~dB.

\begin{figure}[t!]
  \centering
  \includegraphics[width=0.99\columnwidth]{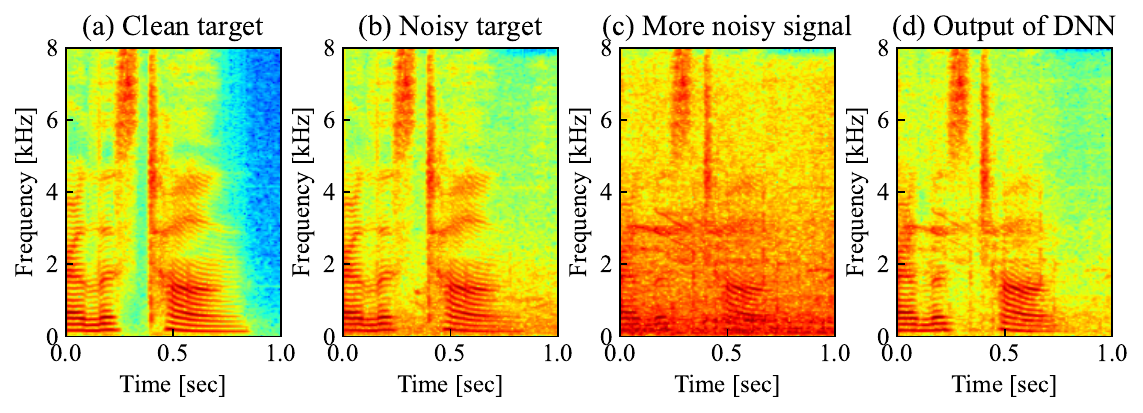} 
  \caption{Spectrograms of the test dataset. (a) Clean target $\bm{s}$, (b) noisy target $\bm{x}$, (c) \textit{more noisy} signal $\bm{y}$, and (d) output of a DNN $f(\bm{y};\theta)$.
  $\bm{n}^{\rm obs}$ and $\bm{n}^{\rm add}$ used for the training were \texttt{CHiME-A} and \texttt{CHiME-B}, respectively.}
\label{fig:spectrogram}
\end{figure}

Table~\ref{tbl:quality} shows the speech quality of 1,000 utterances of the noisy targets $\bm{x}$, the \textit{more noisy} signals $\bm{y}$, and the corresponding output signals $f(\bm{y}; \theta)$, for both the training and test datasets.
Considering that the SI-SDR, PESQ, and STOI of clean speech are $\infty$~dB, $4.64$, and $1.00$, respectively, the output signals are closer in quality to the noisy targets than the clean target signals.
Figure~\ref{fig:spectrogram} provides examples of spectrograms from the test dataset, further illustrating that the output signal is better interpreted as the estimate of the noisy target rather than the clean target signal.

\begin{table}[t]
  \caption{Quality of singles processed in NyTT.
  $\bm{n}^{\mathrm{obs}}$ and $\bm{n}^{\mathrm{add}}$ were \texttt{CHiME-A} and \texttt{CHiME-B}, respectively.}
  \begin{center}
    \scalebox{0.75}{
  \begin{tabular}{c|c|ccc}
  \toprule
  Data & Metric & Noisy target & More noisy & Output \\
  \midrule
  \multirow{3}{*}{Train set} & SI-SDR & \textbf{7.33} & -2.13 & 6.16 \\
   & PESQ & 1.35 & 1.07 & \textbf{1.37} \\
   & STOI & \textbf{0.838} & 0.666 & 0.776 \\
  \midrule
  \multirow{3}{*}{Test set} & SI-SDR & \textbf{9.67} & -1.58 & 7.76 \\
  & PESQ & \textbf{1.48} & 1.08 & 1.47 \\
  & STOI & \textbf{0.874} & 0.696 & 0.811 \\
  \bottomrule
  \end{tabular}
  }
  \end{center}
  \label{tbl:quality}
  \end{table}

\subsubsection{Evaluation of NyTT with loss functions that do not satisfy the conditions of Noise2Noise}
\label{sec:validity2}
If NyTT is Noise2Noise, $\bm{n}^{\rm obs}$ must have a zero-mean distribution, and therefore, the MSE of the loss function should be calculated in the time domain.
To analyze the significance of this assumption in NyTT, we compared the performance of NyTT with MSE in the time domain (\texttt{Time}) and that with MSE in the amplitude spectrogram domain (\texttt{Spec}).
If NyTT strictly adheres to the Noise2Noise framework, \texttt{Spec} should not be able to perform TSE, as the zero-mean distribution for $\bm{n}^{\rm obs}$ cannot be satisfied in the amplitude spectrogram domain.
In this experiment, we estimated real-valued T--F masks for the amplitude spectrograms and transformed the spectrograms to the time-domain signals using the phase of the unprocessed noisy signals.

Table~\ref{tbl:loss} shows the evaluation results, demonstrating that \texttt{Spec} can improve speech quality, despite the lack of the zero-mean assumption for $\bm{n}^{\rm obs}$.
The results in Secs.~\ref{sec:validity1} and \ref{sec:validity2} indicate that NyTT achieves TSE by reducing the noise component corresponding to $\bm{n}^{\rm add}$ rather than the Noise2Noise framework.
Therefore, the zero-mean distribution assumption for $\bm{n}^{\rm obs}$ and the use of the MSE loss function are unnecessary, making NyTT a more flexible training strategy (this conclusion is also supported by the experimental results in Secs.~\ref{sec:dereverb} and \ref{sec:declip}).

\begin{table}[t]
  \caption{Comparison of loss functions. $\bm{n}^{\mathrm{obs}}$ and $\bm{n}^{\mathrm{add}}$ were \texttt{CHiME-A} and \texttt{CHiME-B}, respectively.}
  \begin{center}
    \scalebox{0.75}{
  \begin{tabular}{l|ccc}
  \toprule
   & Unprocessed & \texttt{Time} & \texttt{Spec} \\
  \midrule
  SI-SDR & 9.67 & \textbf{15.89} & 14.94 \\
  PESQ & 1.48 & \textbf{2.33} & 2.10 \\
  STOI & 0.874 & \textbf{0.928} & 0.923 \\
  \bottomrule
  \end{tabular}
    }
  \end{center}
  \label{tbl:loss}
  \end{table}

\subsection{Effectiveness of IterNyTT}
\label{sec:iternytt}
To verify the effectiveness of IterNyTT, we evaluated its performance over five iterations.
In this experiment, $\bm{n}^{\rm obs}$ and $\bm{n}^{\rm add}$ were \texttt{CHiME-A} and \texttt{CHiME-B}, respectively.

\begin{figure}[t!]
  \centering
  \includegraphics[width=\columnwidth]{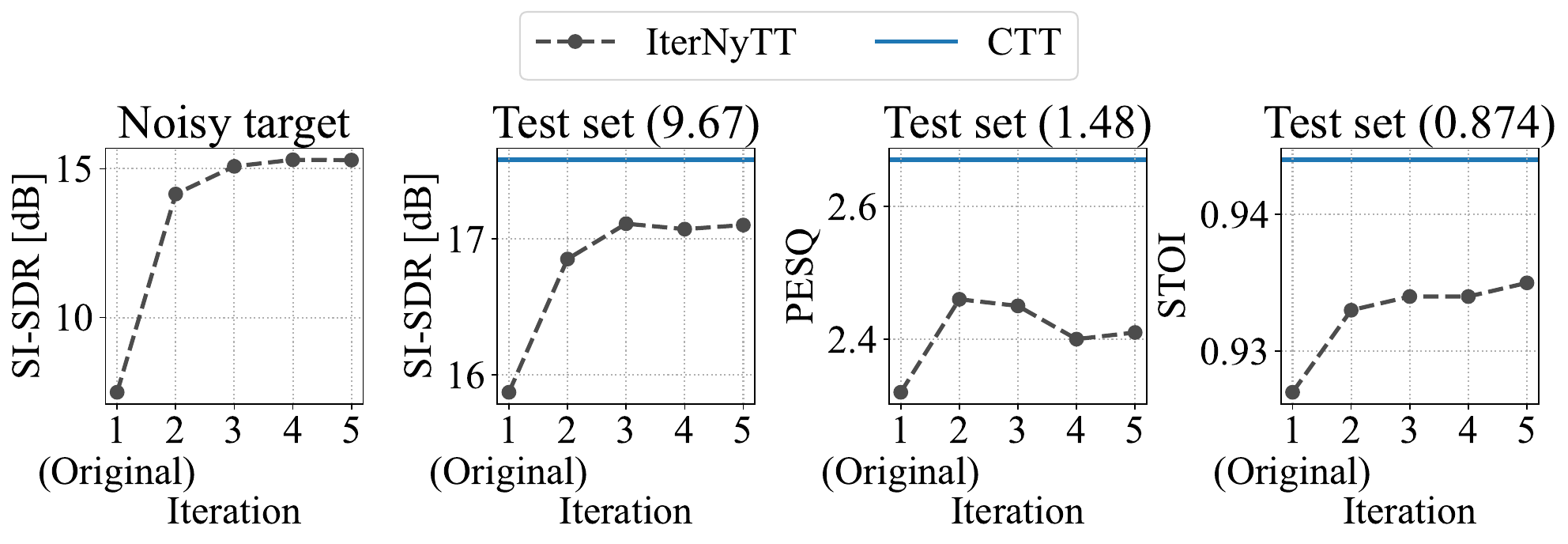} 
  \caption{
    Changes in SI-SDR of the target signals and evaluation results for the test dataset through IterNyTT.
    The first iteration of IterNyTT is equivalent to the original NyTT.
    Values in parentheses indicate the evaluation results of unprocessed input signals.
    $\bm{n}^{\rm obs}$ and $\bm{n}^{\rm add}$ were \texttt{CHiME-A} and \texttt{CHiME-B}, respectively.
    }
\label{fig:iternytt}
\end{figure}

Figure~\ref{fig:iternytt} illustrates the SI-SDR of the noisy targets, along with the SI-SDR, PESQ, and STOI of the processed results for the test dataset at each iteration of IterNyTT.
The figure shows that IterNyTT improves the quality of the noisy targets, and thus, the performance on the test dataset approaches that of CTT as the number of iterations increases.
Additionally, we observe that the SI-SDR of the noisy target is significantly improved at the second iteration and the performance on the test dataset is also improved at that time.

\subsection{Effects of noise mismatches}
\label{sec:noise}
%we conducted three experiments: (1) effects of mismatch on the performance, (2) difference in the impact of $\mathrm{SNR}_{\bm{x}}$ with and without mismatch between $\bm{n}^{\rm obs}$ and $\bm{n}^{\rm test}$, and (3) difference in the impact of $\mathrm{SNR}_{\bm{y}}$ with and without mismatch between $\bm{n}^{\rm obs}$ and $\bm{n}^{\rm test}$.
To investigate the effects of mismatches between noise signals in NyTT (i.e., $\bm{n}^{\rm obs}$, $\bm{n}^{\rm add}$, and $\bm{n}^{\rm test}$),
we simulated mismatched conditions using \texttt{CHiME-A}, \texttt{DEMAND-A}, and \texttt{DCASE} as $\bm{n}^{\rm obs}$, and \texttt{CHiME-B} and \texttt{DEMAND-B} as $\bm{n}^{\rm add}$.

\begin{figure}[t!]
  \centering
  \includegraphics[width=0.99\columnwidth]{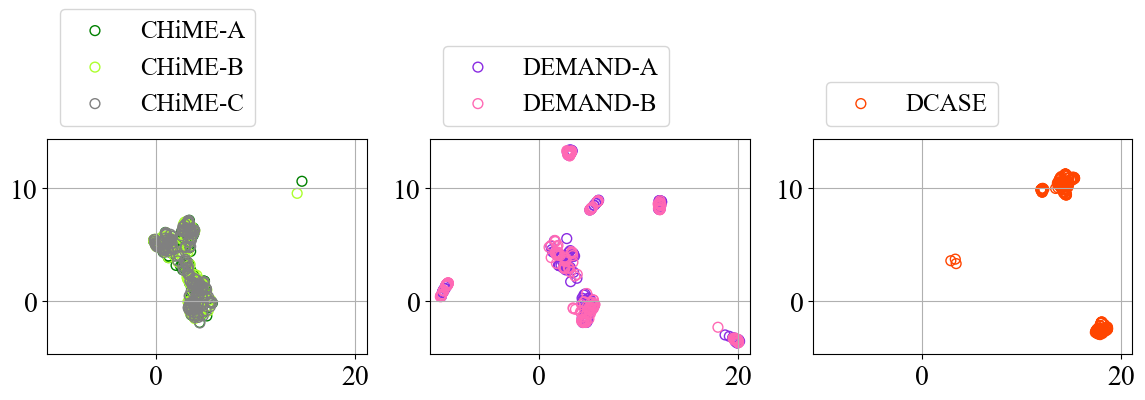} 
  \caption{Distribution of each noise dataset. 
  Although UMAP features were calculated using all noise datasets, they were separately plotted for visibility.
  All figures have the same axes.}
\label{fig:noise}
\end{figure}

To clarify the noise mismatches, we visualize the distribution of each noise dataset in Fig.~\ref{fig:noise}.
The plots in Fig.~\ref{fig:noise} were created by extracting features using a pre-trained audio event classification model, VGGish~\cite{hershey2017cnn}, and projecting them into a two-dimensional space using UMAP~\cite{mcinnes2018umap}.
For visibility, we randomly selected 200 samples of \SI{2}{s} noise signals from each dataset.
The figure shows that each noise dataset forms a distinct cluster, indicating that they have different characteristics.
We can also see that \texttt{DCASE} has particularly different characteristics from the other datasets.

\subsubsection{Effects of mismatches on the performance}
\label{sec:noise1}
We investigated the effects of mismatches between noise signals on the performance of CTT, NyTT, and IterNyTT.
In this experiment, we set the number of iterations for IterNyTT to three.

Table~\ref{tbl:noise} shows the evaluation results of CTT, NyTT, and IterNyTT for each combination of $\bm{n}^{\rm obs}$ and $\bm{n}^{\rm add}$.
The table also includes the evaluation results of IterNyTT using different noise datasets for the first and second iterations (training for the TSE of the noisy targets) and the third iteration (training for the TSE of the test dataset).
We analyze these results from three perspectives: mismatches between a) $\bm{n}^{\rm add}$ and $\bm{n}^{\rm test}$, b) $\bm{n}^{\rm obs}$ and $\bm{n}^{\rm test}$, and c) $\bm{n}^{\rm obs}$ and $\bm{n}^{\rm add}$, where $\bm{n}^{\rm test}$ was \texttt{CHiME-C}.

First, we focus on the impact of the mismatch between $\bm{n}^{\rm add}$ and $\bm{n}^{\rm test}$ on the performance of NyTT.
%For example, NyTT achieved SI-SDR of 15.99~dB when $(\bm{n}^{\rm obs}, \bm{n}^{\rm add})$ is (\texttt{CHiME-A}, \texttt{CHiME-B}), and 10.78~dB when $(\bm{n}^{\rm obs}, \bm{n}^{\rm add})$ is (\texttt{CHiME-A}, \texttt{DEMAND-B}).
For example, we analyze the performance of NyTT when $\bm{n}^{\rm obs}$ is \texttt{CHiME-A}.
In this case, NyTT achieves SI-SDRs of 15.87~dB when $\bm{n}^{\rm add}$ is \texttt{CHiME-B} and 10.27~dB when $\bm{n}^{\rm add}$ is \texttt{DEMAND-B}.
Similarly, even when $\bm{n}^{\rm obs}$ is \texttt{DEMAND-A} or \texttt{DCASE}, and even when the metric is PESQ or STOI, we can consistently see that NyTT performs better when there is no mismatch between $\bm{n}^{\rm add}$ and $\bm{n}^{\rm test}$, as in CTT.
% for NyTT with other $\bm{n}^{\rm obs}$ and CTT, 

Second, we focus on the impact of the mismatch between $\bm{n}^{\rm obs}$ and $\bm{n}^{\rm test}$ on the performance of NyTT.
For example, we analyze the performance of NyTT when $\bm{n}^{\rm add}$ is \texttt{DEMAND-B}.
In this case, NyTT achieves SI-SDRs of 10.27~dB when $\bm{n}^{\rm obs}$ is \texttt{CHiME-A}, 13.56~dB when $\bm{n}^{\rm obs}$ is \texttt{DEMAND-A}, and 14.20~dB when $\bm{n}^{\rm obs}$ is \texttt{DCASE}.
Similarly, even when $\bm{n}^{\rm add}$ is \texttt{CHiME-B}, and even when the metric is PESQ or STOI, we can consistently see that NyTT performs better when there is a mismatch between $\bm{n}^{\rm obs}$ and $\bm{n}^{\rm test}$.
In particular, we can see that NyTT achieves its best performance when $\bm{n}^{\rm obs}$ is \texttt{DCASE}, which has distinctly different characteristics from \texttt{CHiME-C} of $\bm{n}^{\rm test}$, as shown in Fig.~\ref{fig:noise}.
%$\bm{n}^{\rm obs}$ is \texttt{DEMAND-A} or \texttt{DCASE} compared to when $\bm{n}^{\rm obs}$ is \texttt{CHiME-A}.
%when $\bm{n}^{\rm add}$ is \texttt{CHiME-B}, 
Since the DNN is trained to include $\bm{n}^{\rm obs}$ in the output signals in NyTT, it is expected that noise will remain in the output signals when there is no mismatch between $\bm{n}^{\rm test}$ and $\bm{n}^{\rm obs}$.

Third, we focus on the impact of the mismatch between $\bm{n}^{\rm obs}$ and $\bm{n}^{\rm add}$ on the performance of IterNyTT.
For example, we analyze the performance of IterNyTT when $\bm{n}^{\rm obs}$ is \texttt{CHiME-A}.
In this case, IterNyTT achieves SI-SDRs of 17.11~dB when $\bm{n}^{\rm add}$ is \texttt{CHiME-B} and 14.64~dB when $\bm{n}^{\rm add}$ is (\texttt{DEMAND-B}, \texttt{CHiME-B}).
%When $\bm{n}^{\rm add}$ is (\texttt{DEMAND-B}, \texttt{CHiME-B}), the mismatch between $\bm{n}^{\rm obs}$(=\texttt{CHiME-A}) and $\bm{n}^{\rm add}$(=\texttt{DEMAND-B}) prevents IterNyTT from effectively removing $\bm{n}^{\rm obs}$ from the noisy targets in the first and second iterations, resulting in no performance improvement on the test dataset in the third iteration.
When $\bm{n}^{\rm add}$ is (\texttt{DEMAND-B}, \texttt{CHiME-B}), in the first and second iterations, the mismatch between $\bm{n}^{\rm obs}$(=\texttt{CHiME-A}) and $\bm{n}^{\rm add}$(=\texttt{DEMAND-B}) prevents IterNyTT from effectively removing $\bm{n}^{\rm obs}$ from the noisy targets.
Thus, there is no performance improvement on the test dataset in the third iteration.
Additionally, IterNyTT achieves a higher SI-SDR when $\bm{n}^{\rm add}$ is (\texttt{CHiME-B}, \texttt{DEMAND-B}) (10.67 dB) than when $\bm{n}^{\rm add}$ is \texttt{DEMAND-B} (9.80 dB).
Moreover, IterNyTT shows little performance improvement when $\bm{n}^{\rm obs}$ is \texttt{DCASE}.
Similarly, even when $\bm{n}^{\rm obs}$ is \texttt{DEMAND-A}, and even when the metric is PESQ or STOI, we can consistently see that IterNyTT performs better when using $\bm{n}^{\rm add}$ matched with $\bm{n}^{\rm obs}$ in the first and second iterations.
The impact of the mismatch between $\bm{n}^{\rm obs}$ and $\bm{n}^{\rm add}$ on IterNyTT is consistent with that of the mismatch between $\bm{n}^{\rm add}$ and $\bm{n}^{\rm test}$ on NyTT.
%there is no mismatch between $\bm{n}^{\rm obs}$ and $\bm{n}^{\rm add}$.
% Additionally, we observe that even when there is no mismatch between $\bm{n}^{\rm obs}$ and $\bm{n}^{\rm test}$, IterNyTT can recover the performance by using proper $\bm{n}^{\rm add}$ (e.g., when $\bm{n}^{\rm obs}$ and $\bm{n}^{\rm test}$ are \texttt{CHiME-A} and \texttt{CHiME-B}, respectively).
%when $\bm{n}^{\rm obs}$ is \texttt{DEMAND-A}, 
% Particularly, when $\bm{n}^{\rm obs}$ is \texttt{CHiME-A} and there is no mismatch between $\bm{n}^{\rm obs}$ and $\bm{n}^{\rm add}$, the performance of the original NyTT is degraded; however, IterNyTT successfully recovers and achieves significant improvement by using \texttt{CHiME-B} as $\bm{n}^{\rm add}$.

\begin{figure}[t!]
  \centering
  \includegraphics[width=0.8\columnwidth]{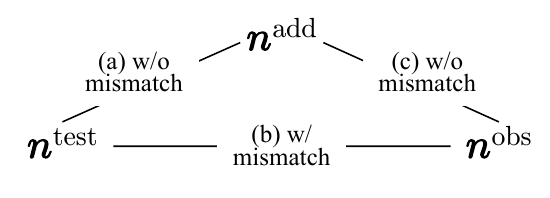} 
  \caption{Desirable condition in NyTT framework.}
\label{fig:desirable}
\end{figure}

Summing up the above results, we derive the desirable condition for the NyTT framework, as shown in Fig.~\ref{fig:desirable}.
a) NyTT achieves high performance when there is no mismatch between $\bm{n}^{\rm add}$ and $\bm{n}^{\rm test}$, as in CTT, 
b) NyTT achieves high performance when there is a mismatch between $\bm{n}^{\rm obs}$ and $\bm{n}^{\rm test}$, and 
c) IterNyTT improves the performance when there is no mismatch between $\bm{n}^{\rm obs}$ and $\bm{n}^{\rm add}$.
Additionally, these results are consistent with the interpretation of NyTT in Sec.~\ref{sec:validity}.

\begin{table*}[t]
\caption{Evaluation results on the test dataset.
The SI-SDR, PESQ, and STOI of the unprocessed noisy signals were 10.27~dB, 1.48, and 0.874, respectively.
$\bm{n}^{\rm test}$ was \texttt{CHiME-C}.
When IterNyTT used two noise datasets as $\bm{n}^{\rm add}$, the first noise dataset in parentheses was used in the first and second iterations of IterNyTT, whereas the second noise dataset in parentheses was used in the third iteration of IterNyTT.
In CTT, the choice of $\bm{n}^{\rm obs}$ is not involved.}

\begin{center}
\resizebox{\columnwidth}{!}{
\begin{tabular}{cc|ccc|ccc|ccc}
% \toprule
% $\bm{n}^{\rm obs}$ & $\bm{n}^{\rm add}$ & CTT & NyTT & IterNyTT\\
% \midrule
% \multirow[c]{4}{*}{\texttt{CHiME-A}} & \texttt{CHiME-B} & \textbf{17.66} & 15.99 & 17.20 \\
%  & \texttt{DEMAND-B} & \textbf{15.22} & 10.78 & 10.38 \\
%  & (\texttt{CHiME-B}, \texttt{DEMAND-B}) & - & - & 11.18\\
%  & (\texttt{DEMAND-B}, \texttt{CHiME-B}) & - & - & 14.81\\
 
% \midrule
% \multirow[c]{4}{*}{\texttt{DEMAND-A}} & \texttt{CHiME-B} & \textbf{17.66} & 16.69 & 17.29 \\
%  & \texttt{DEMAND-B} & \textbf{15.22} & 13.80 & 14.34 \\
%  & (\texttt{CHiME-B}, \texttt{DEMAND-B}) & - & - & 14.09\\
% & (\texttt{DEMAND-B}, \texttt{CHiME-B}) & - & - & 17.42\\

% \midrule
% \multirow[c]{2}{*}{\texttt{DCASE}} & \texttt{CHiME-B} & \textbf{17.66} & 16.85 & 17.16 \\
%  & \texttt{DEMAND-B} & \textbf{15.22} & 14.42 & 13.81 \\
% % \multirow[c]{2}{*}{\texttt{CHiME-A}} & (\texttt{CHiME-B}, \texttt{DEMAND-B}) & - & - & 11.18\\
% %  & (\texttt{DEMAND-B}, \texttt{CHiME-B}) & - & - & 14.81\\

% % \multirow[c]{2}{*}{\texttt{DEMAND-A}} & (\texttt{CHiME-B}, \texttt{DEMAND-B}) & - & - & 14.09\\
% % & (\texttt{DEMAND-B}, \texttt{CHiME-B}) & - & - & 17.42\\
  \toprule
  \multirow[c]{2}{*}{$\bm{n}^{\rm obs}$} & \multirow[c]{2}{*}{$\bm{n}^{\rm add}$} & \multicolumn{3}{c|}{SISDR} & \multicolumn{3}{c|}{PESQ} & \multicolumn{3}{c}{STOI} \\
   &  & CTT & NyTT & IterNyTT & CTT & NyTT & IterNyTT & CTT & NyTT & IterNyTT \\
  \midrule
  \multirow[c]{4}{*}{\texttt{CHiME-A}} & \texttt{CHiME-B} & \textbf{17.58} & 15.87 & 17.11 & \textbf{2.67} & 2.32 & 2.45 & \textbf{0.944} & 0.927 & 0.934 \\
   & \texttt{DEMAND-B} & \textbf{15.04} & 10.27 & 9.80 & \textbf{2.16} & 1.57 & 1.51 & \textbf{0.926} & 0.882 & 0.877 \\
   & (\texttt{CHiME-B}, \texttt{DEMAND-B}) & - & - & 10.67 & - & - & 1.59 & - & - & 0.881 \\
   & (\texttt{DEMAND-B}, \texttt{CHiME-B}) & - & - & 14.64 & - & - & 1.96 & - & - & 0.915 \\
   \midrule
  \multirow[c]{4}{*}{\texttt{DEMAND-A}} & \texttt{CHiME-B} & \textbf{17.58} & 16.59 & 17.20 & \textbf{2.67} & 2.53 & 2.48 & \textbf{0.944} & 0.936 & 0.937 \\
   & \texttt{DEMAND-B} & \textbf{15.04} & 13.56 & 14.11 & \textbf{2.16} & 1.98 & 2.06 & \textbf{0.926} & 0.905 & 0.917 \\
   & (\texttt{CHiME-B}, \texttt{DEMAND-B}) & - & - & 13.85 & - & - & 1.94 & - & - & 0.903 \\
   & (\texttt{DEMAND-B}, \texttt{CHiME-B}) & - & - & 17.33 & - & - & 2.56 & - & - & 0.940 \\
   \midrule
  \multirow[c]{2}{*}{\texttt{DCASE}} & \texttt{CHiME-B} & \textbf{17.58} & 16.75 & 17.06 & \textbf{2.67} & 2.47 & 2.45 & \textbf{0.944} & 0.937 & 0.937 \\
   & \texttt{DEMAND-B} & \textbf{15.04} & 14.20 & 13.54 & \textbf{2.16} & 2.06 & 1.91 & \textbf{0.926} & 0.914 & 0.908 \\
  \bottomrule
  \end{tabular}
}
\end{center}
\label{tbl:noise}
\end{table*}

\subsubsection{Difference in the impact of $\mathrm{SNR}_{\textbf{x}}$ with and without the mismatch between $\textbf{n}^{\rm obs}$ and $\textbf{n}^{\rm test}$}
\label{sec:noise2}
%NyTT has a property in which the performance degrades when $\mathrm{SNR}_{\bm{x}}$ is low~\cite{Fujimura_2021}.
NyTT experiences performance degradation when $\mathrm{SNR}{\bm{x}}$ is low~\cite{Fujimura_2021}.
However, the impact of $\mathrm{SNR}{\bm{x}}$ on performance is expected to vary depending on the mismatch between $\bm{n}^{\rm obs}$ and $\bm{n}^{\rm test}$.
%For example, when there is no mismatch between $\bm{n}^{\rm obs}$ and $\bm{n}^{\rm test}$, a higher $\mathrm{SNR}{\bm{x}}$ is likely required.
To investigate this, we evaluated the performance of NyTT using \texttt{CHiME-A}, \texttt{DEMAND-A}, and \texttt{DCASE} as $\bm{n}^{\rm obs}$, and \texttt{CHiME-B} as $\bm{n}^{\rm add}$, and by adjusting $\mathrm{SNR}_{\bm{x}}$ to -5, 0, 5, 10, 15, and 20~dB.
Additionally, we evaluated the performance of CTT for the case where $\mathrm{SNR}_{\bm{x}}$ is $\infty$ dB.
In this experiment, $\mathrm{SNR}_{\bm{y}}$ was set to range from -5 to 5~dB for both CTT and NyTT. % to ensure a fair comparison.

\begin{figure}[t!]
  \centering
  \includegraphics[width=0.98\columnwidth]{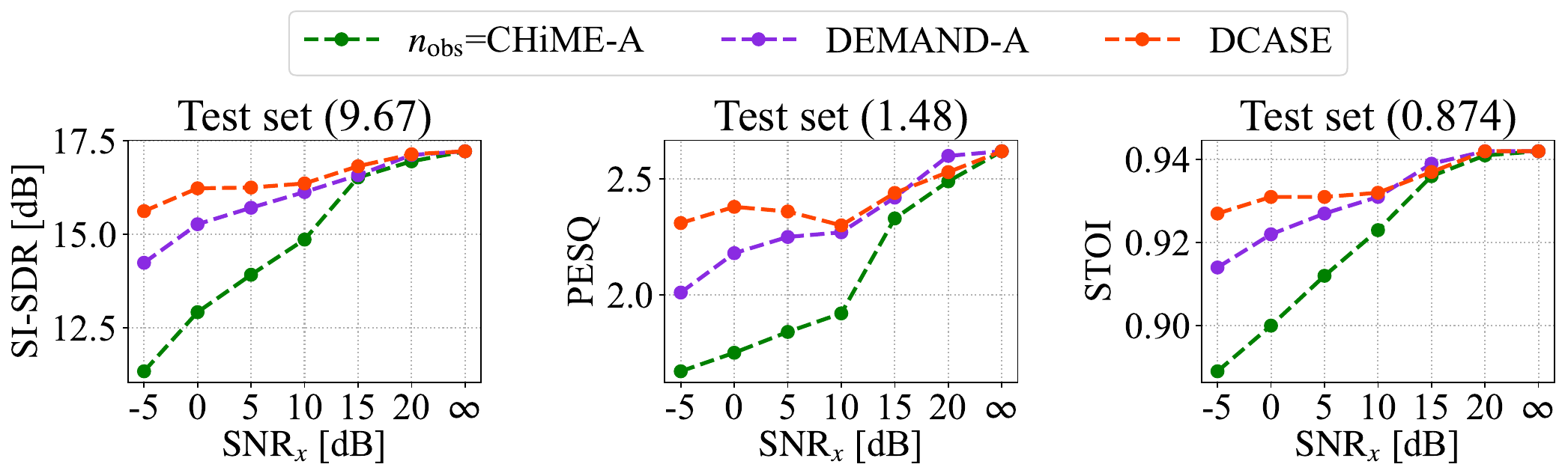} 
  \caption{Relationship between $\mathrm{SNR}_{\bm{x}}$ and the evaluation results of NyTT. 
  Values in parentheses indicate the evaluation results of unprocessed input signals.
  $\bm{n}^{\rm add}$ was \texttt{CHiME-B} and $\mathrm{SNR}_{\bm{y}}$ ranged from -5 to 5 dB.}
\label{fig:snr_x}
\end{figure}

Figure~\ref{fig:snr_x} shows the SI-SDR, PESQ, and STOI of the processed results for the test dataset at each $\mathrm{SNR}_{\bm{x}}$.
Overall, the performance of NyTT degrades as $\mathrm{SNR}_{\bm{x}}$ decreases.
The impact depends on mismatches; significant degradation occurs when $\bm{n}^{\rm obs}$ is \texttt{CHiME-A}, which has no mismatch with $\bm{n}^{\rm test}$.
In contrast, when there is a mismatch, the performance remains relatively high even at a low $\mathrm{SNR}_{\bm{x}}$.
From the results of this experiment, we confirmed that the impact of $\mathrm{SNR}_{\bm{x}}$ significantly depends on the mismatch between $\bm{n}^{\rm obs}$ and $\bm{n}^{\rm test}$.
%and the impact of $\mathrm{SNR}_{\bm{x}}$ is significant when $\bm{n}^{\rm obs}$ is \texttt{CHiME-A} which has no mismatch with $\bm{n}^{\rm test}$.

\subsubsection{Difference in the impact of $\mathrm{SNR}_{\textbf{y}}$ with and without the mismatch between $\textbf{n}^{\rm obs}$ and $\textbf{n}^{\rm test}$}
\label{sec:noise3}
Considering that NyTT trains a DNN to estimate noisy targets by removing $\bm{n}^{\rm add}$ from \textit{more noisy} signals, $\mathrm{SNR}_{\bm{y}}$ also affects the performance.
For example, when there is no mismatch between $\bm{n}^{\rm obs}$ and $\bm{n}^{\rm test}$, and $\mathrm{SNR}_{\bm{y}}$ is high, the effect of reducing noise is less significant than the adverse effect of the residual noise in the output signal.
To investigate this, we evaluated the performance of NyTT using \texttt{CHiME-A} and \texttt{DCASE} as $\bm{n}^{\rm obs}$, and \texttt{CHiME-B} as $\bm{n}^{\rm add}$, and by varying the $\mathrm{SNR}_{\bm{y}}$ range to $[-10, -5)$, $[-5, 0)$, $[0, 5)$, $[5, 10)$, and $[10, 15)$ dB.
We also evaluated the performance of CTT.
In this experiment, we set $\mathrm{SNR}_{\bm{x}}$ to 5 dB for NyTT.

\begin{figure}[t!]
  \centering
  \includegraphics[width=0.98\columnwidth]{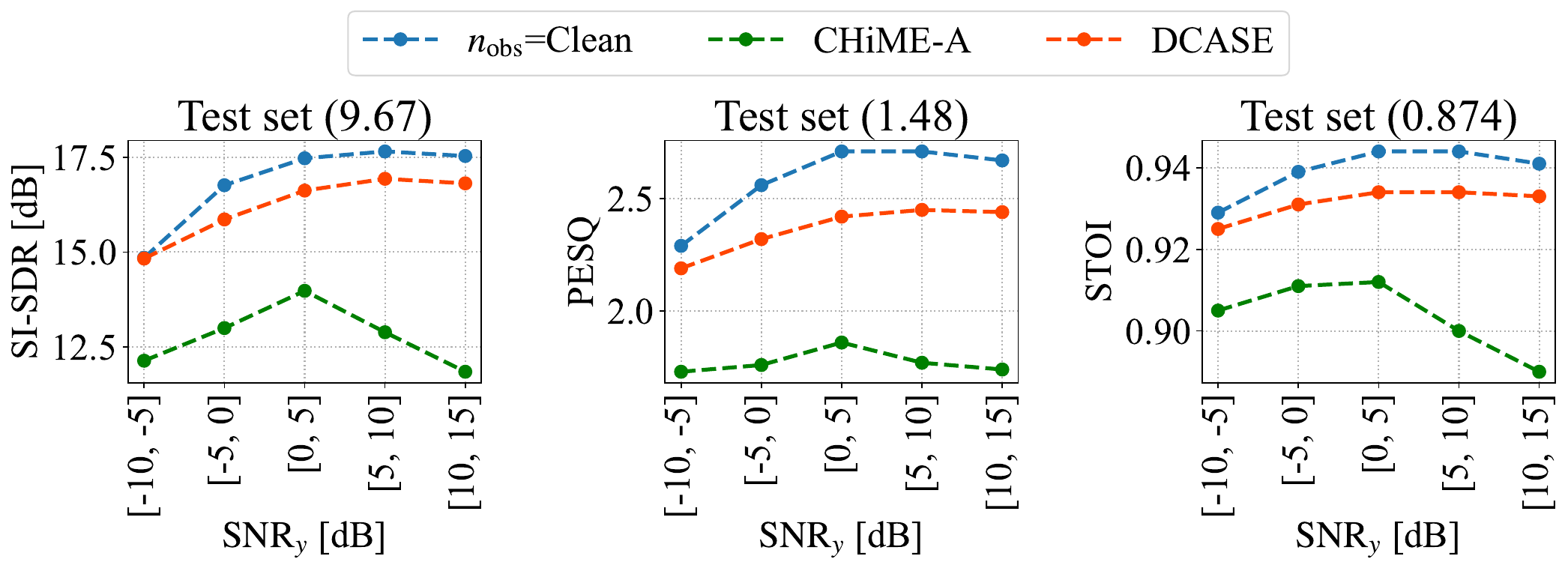} 
  \caption{Relationship between $\mathrm{SNR}_{\bm{y}}$ and the evaluation results of NyTT. 
  Values in parentheses indicate the evaluation results of unprocessed input signals.
  $\bm{n}^{\rm add}$ was \texttt{CHiME-B} and $\mathrm{SNR}_{\bm{x}}$ was 5~dB.}
\label{fig:snr_y}
\end{figure}

Figure~\ref{fig:snr_y} shows the SI-SDR, PESQ, and STOI of the processed results for the test dataset at each $\mathrm{SNR}_{\bm{y}}$ range, where the performance of CTT varies depending on the mismatch of SNR between the training and test datasets.
When $\bm{n}^{\rm obs}$ is \texttt{DCASE}, which has a mismatch with $\bm{n}^{\rm test}$, the performance of NyTT has a similar tendency to that of CTT.
On the other hand, when $\bm{n}^{\rm obs}$ is \texttt{CHiME-A}, which has no mismatch with $\bm{n}^{\rm test}$, the performance of NyTT degrades when $\mathrm{SNR}_{\bm{y}}$ exceeds 5 dB, and this trend differs from that of CTT.
Thus, when there is no mismatch between $\bm{n}^{\rm obs}$ and $\bm{n}^{\rm test}$, NyTT effectively acquires the TSE feature by setting $\mathrm{SNR}_{\bm{y}}$ to a moderately low level, not too low.

%Progress: 2024/10/15,19:00

\subsection{Effectiveness of utilizing noisy signals in a situation where clean target signals are available}
\label{sec:volume}
In this experiment, we used 100 utterances from LibriSpeech as \texttt{Clean-100} and 900 utterances as \texttt{Clean-900}.
We generated \texttt{Noisy-900} by mixing \texttt{Clean-900} with \texttt{CHiME-A} at an SNR of 5~dB.
We investigated the effectiveness of using \texttt{Noisy-900}.
Moreover, we investigated the effectiveness of using \texttt{EnhNoisy-900}, which was generated by applying TSE to \texttt{Noisy-900} using the CTT model trained on \texttt{Clean-100}.
In this experiment, we used \texttt{CHiME-B} as $\bm{n}^{\rm add}$, and we used 50 clean utterances from LibriSpeech for the validation of both CTT and NyTT.
Note that there was no mismatch between $\bm{n}^{\rm obs}$ and $\bm{n}^{\rm test}$, and $\mathrm{SNR}_{\bm{x}}$ was 5~dB, resulting in a challenging condition for NyTT.
Since the volumes of the speech datasets were different, we carefully trained a DNN for enough epochs, ensuring that the best epoch remained unchanged for the last 300 epochs.

Table~\ref{tbl:volume} shows the evaluation results, demonstrating that the combined use of \texttt{Clean-100} and \texttt{Noisy-900} achieves a higher SI-SDR than using either dataset separately.
Furthermore, the performance of the combined use of \texttt{Clean-100} and \texttt{EnhNoisy-900} approaches that of the ideal situation where both \texttt{Clean-100} and \texttt{Clean-900} are available.
We can also expect that the performance will improve with the use of noisy targets recorded under better conditions (i.e., higher $\mathrm{SNR}_{\bm{x}}$ or $\bm{n}^{\rm obs}$ mismatched with $\bm{n}^{\rm test}$).
These results indicate that leveraging a large number of noisy signals is beneficial, even when a small number of clean target signals are available.

\begin{table}[t]
\caption{Evaluation results of the combined use of the clean and noisy signals. 
$\bm{n}^{\rm obs}$ was \texttt{CHiME-A}, $\bm{n}^{\rm add}$ was \texttt{CHiME-B}, and $\mathrm{SNR}_{\bm{x}}$ was 5 dB.
SI-SDR, PESQ, and STOI of the unprocessed noisy signals were 10.27~dB, 1.48, and 0.874, respectively.
$\bm{n}^{\rm test}$ was \texttt{CHiME-C}. We assume that \texttt{Clean-900} is not available.}
\begin{center}
  \scalebox{0.75}{
\begin{tabular}{c|ccc|c}
  \toprule
  Training dataset & SI-SDR &  PESQ & STOI & Epoch \\
  \midrule
  \texttt{Clean-100} & 13.82 & 2.03 & 0.910 & 3,816 \\
  \texttt{Noisy-900} & 13.68 & 1.88 & 0.906 & 1,302 \\
  \texttt{Clean-100}, \texttt{Noisy-900} & 14.35 & 2.00 & 0.914 & 2,350 \\
  \texttt{Clean-100}, \texttt{EnhNoisy-900} & 16.45 & 2.33 & 0.935 & 11,626 \\
  % enhanced-noisy-900 & 16.46 & 2.30 & 15000 & 14215 \\
  \gray{\texttt{Clean-100}, \texttt{Clean-900}} & \gray{\textbf{17.13}} & \gray{\textbf{2.56}} & \gray{\textbf{0.942}} & \gray{11,593} \\
\bottomrule
\end{tabular}
  }
\end{center}
\label{tbl:volume}
\end{table}

\section{Experimental analysis in the dereverberation task}
\label{sec:dereverb}

\subsection{Setups}
In the experiments, we used RIR simulated by utilizing Pyroomacoustics~\cite{Scheibler2018pyroomacoustics}.
The room width and depth were randomly selected from 4 to 8 m, the height was randomly selected from 2 to 6 m, and the distance between the microphone and the sound source was set to 1 m.
The reverberation time $\mathrm{RT}_{60}$ ranged from 0.20 to \SI{1.10}{s}.
We divided the range into \SI{0.05}{s} segments and generated 170 RIRs for each interval.
The 170 RIRs were split into 80, 80, and 10 samples, and which were used as \texttt{RIR-A}, \texttt{RIR-B}, and \texttt{RIR-C}, respectively.
The total number of \texttt{RIR-A}, \texttt{RIR-B}, and \texttt{RIR-C} were 1,440, 1,440, and 180, respectively.
Each of these three datasets covers the same $\mathrm{RT}_{60}$ range.
\texttt{RIR-A}, \texttt{RIR-B}, and \texttt{RIR-C} were used as $\bm{r}^{\rm obs}$, $\bm{r}^{\rm add}$, and $\bm{r}^{\rm test}$, respectively.
% \texttt{RIR-A} was split based on $\mathrm{RT}_{60}$ and used as $\bm{r}^{\rm obs}$, and \texttt{RIR-B} and \texttt{RIR-C} were used as $\bm{r}^{\rm add}$ and $\bm{r}^{\rm test}$, respectively.
The clean target signals were 10,000 utterances from LibriSpeech and the reverberant target signals were generated by convolving the clean target signals with \texttt{RIR-A}.
The test dataset of reverberant signals was generated by convolving 1,000 utterances from LibriSpeech with \texttt{RIR-C}.
The sampling frequency was 16 kHz.

The DNN was Conv-TasNet~\cite{luo2019conv} implemented in the Asteroid toolkit~\cite{pariente2020asteroid}, and the loss function was SNR.
We trained the DNN for 850 epochs with a mini-batch size of 12, using the Adam optimizer with a fixed learning rate of 0.0001.
For the validation of both CTT and NyTT, we used 50 clean utterances of LibriSpeech and generated reverberant signals using the same $\bm{r}^{\rm add}$ as in the training.
As the metrics, we used speech-to-reverberation modulation energy ratio~(SRMR)~\cite{falk2010non} in addition to SI-SDR, PESQ, and STOI.

\subsection{Effectiveness of NyTT in the dereverberation task}
We conducted experimental evaluations of NyTT in the dereverberation task.
Table~\ref{tbl:dereverb} shows the evaluation results under different $\mathrm{RT}{60}$ conditions of $\bm{r}^{\rm obs}$.
First, we observe that NyTT achieves higher scores than the unprocessed signals in most cases, demonstrating its capability in this task.
This result also demonstrates that NyTT is not Noise2Noise, since the degradation is not even caused by additive noise.
Additionally, we can see that the performance improves with shorter $\mathrm{RT}{60}$ values and degrades with longer $\mathrm{RT}{60}$ values.
This trend is consistent with the results of the denoising task, where higher quality ($\mathrm{SNR}_{\bm{x}}$) leads to better performance.

\begin{table}[t]
  \caption{Evaluation results of CTT and NyTT in the dereverberation task.}
  \begin{center}
    \scalebox{0.75}{
    \begin{tabular}{cc|ccccccc}
      \toprule
      Method & $\mathrm{RT}_{60}$ of $\bm{r}^{\rm obs}$ [sec]& SI-SDR & PESQ & STOI & SRMR \\
      \midrule
      Unprocessed & - & -5.32 & 1.59 & 0.834 & 4.84 \\
      CTT & 0.0 & \textbf{3.83} & \textbf{2.23} & \textbf{0.918} & \textbf{8.81} \\
      NyTT & $[0.20,0.50)$ & 1.69 & 1.95 & 0.902 & 7.59 \\
      NyTT & $[0.50,0.80)$ & 0.51 & 1.74 & 0.882 & 6.01 \\
      NyTT & $[0.80, 1.10)$ & -1.39 & 1.57 & 0.840 & 4.85 \\
      \bottomrule
      \end{tabular}
    }
  \end{center}
  \label{tbl:dereverb}
\end{table}

% To confirm whether we can apply NyTT to the dereverberation task, we conducted experimental evaluation of CTT and NyTT.
% Table~\ref{tbl:dereverb} shows the evaluation results in the different $\bm{r}^{\rm obs}$  $\mathrm{RT}_{60}$ conditions.
% First, we can see that NyTT with any $\mathrm{RT}_{60}$ of $\bm{r}^{\rm obs}$ achieves SI-SDR and PESQ higher than the unprocessed signals.
% Furthermore, the performance will be improved with shorter $\mathrm{RT}_{60}$ of $\bm{r}^{\rm obs}$.
% Furthermore, as in the denoising task, the performance of NyTT is degraded when the $\mathrm{RT}_{60}$ of $\bm{r}^{\rm obs}$ is longer and the quality of the noisy target is low.
% These results indicate that NyTT similarly works for various distortion.
%we can say that NyTT is a general training strategy for reconstruction tasks.

\subsection{Effectiveness of IterNyTT in the dereverberation task}
\begin{figure}[t!]
  \centering
  \includegraphics[width=0.98\columnwidth]{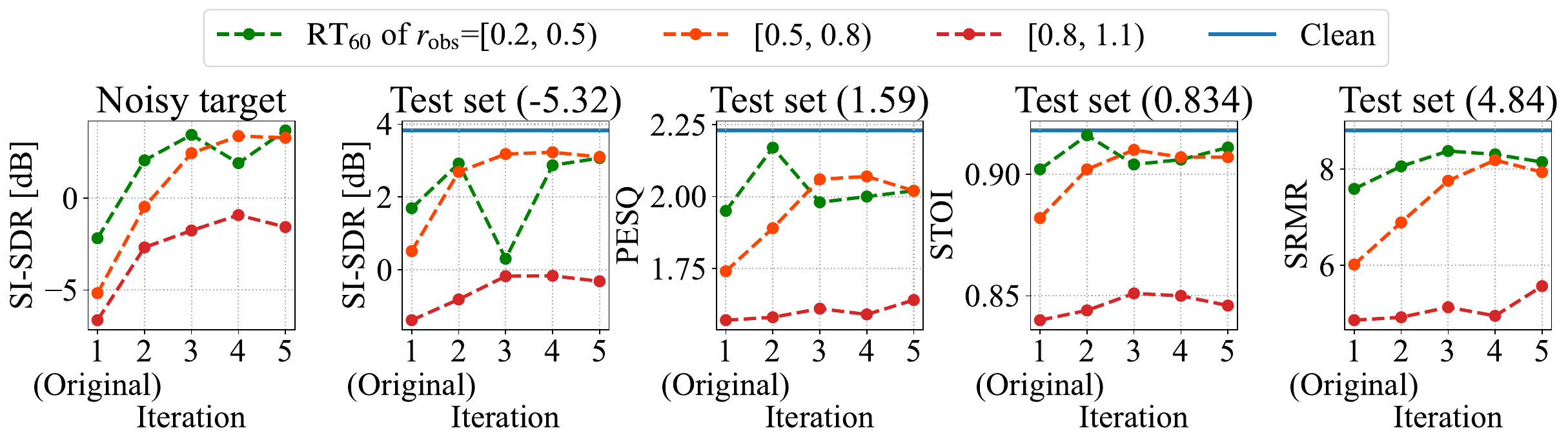} 
  \caption{Changes in SI-SDR of the target signals and evaluation results on the test dataset through IterNyTT under different $\mathrm{RT}_{60}$ conditions of $\bm{r}^{\rm obs}$.
  The first iteration of IterNyTT is equivalent to the original NyTT.
  Values in parentheses indicate the evaluation results of unprocessed input signals.}
\label{fig:dereverb_iternytt}
\end{figure}

To verify the effectiveness of IterNyTT in the dereverberation task, we evaluated the performance over five iterations under different $\mathrm{RT}_{60}$ conditions for $\bm{r}^{\rm obs}$.
Figure~\ref{fig:dereverb_iternytt} illustrates the SI-SDR of the reverberant targets, along with the SI-SDR, PESQ, STOI, and SRMR of the processed results for the test dataset at each iteration of IterNyTT.
From this figure, we observe an overall trend of improved performance of IterNyTT.
When $\mathrm{RT}_{60}$ of $\bm{r}^{\rm obs}$ was $[0.8, 1.1)$, IterNyTT in the first iteration does not perform well, and the performance is not improved in the subsequent iterations.
We can also see that IterNyTT works stably when the $\mathrm{RT}_{60}$ of $\bm{r}^{\rm obs}$ is $[0.5, 0.8)$, whereas it becomes unstable when the $\mathrm{RT}_{60}$ of $\bm{r}^{\rm obs}$ is $[0.2, 0.5)$.
Although there are cases where IterNyTT does not work well, especially when the target signals are of very low quality, we can conclude that IterNyTT is generally effective even in the dereverberation task.
%, demonstrating that IterNyTT is also effective for dereverberation.
%In addition, we can see that NyTT achieves performance comparable to CTT using the noisy target having an SI-SDR of approximately 5~dB.

\section{Experimental analysis in the declipping task}
\label{sec:declip}

\subsection{Setups}
In the experiments, the clean target signals were 10,000 utterances of LibriSpeech, and we generated the clipped target signals $\bm{x}$ by clipping them with an $\mathrm{SNR}_{\bm{x}}$ of 3, 7, or 15~dB.
The clipped signals of the test dataset were generated by clipping 1,000 utterances of LibriSpeech with the SNR randomly selected from 1, 3, 7, and 15~dB.
During the training, for both CTT and NyTT, the clipping threshold was determined from the $\mathrm{SNR}_{\bm{y}}$ randomly selected from 1 to 9~dB.

The DNN was a causal Demucs~\cite{yi2024ddd,defossez20_interspeech}, and the loss function was a weighted sum of the L1 waveform and multi-resolution STFT losses, as in \cite{kwon2024speech,yi2024ddd}.
The weights for the L1 waveform and multi-resolution STFT losses were set to 10 and 0.1, respectively.
We trained the DNN for 400 epochs with a mini-batch size of 12, using the Adam optimizer with a fixed learning rate of 0.0001.
For the validation, we used 50 clean utterances of LibriSpeech and generated clipped signals with the SNR randomly selected from 1, 3, 7, and 15~dB.
As the metrics, we used SI-SDR, PESQ, and STOI.

\begin{table}[t]
  \caption{Evaluation results of CTT and NyTT in the declipping task.}
  \begin{center}
    \scalebox{0.75}{
    \begin{tabular}{c|cccccc}
        \toprule
        Method & $\mathrm{SNR}_{\bm{x}}$ [dB] & SI-SDR & PESQ & STOI \\
        \midrule
        Unprocessed & - & 6.41 & 1.89 & 0.866 \\
        CTT & $\infty$ & \textbf{16.59} & \textbf{3.53} & \textbf{0.965} \\
        NyTT & 15 & 15.27 & 3.21 & 0.959 \\
        NyTT & 7 & 12.44 & 2.65 & 0.941 \\
        NyTT & 3 & 10.09 & 2.28 & 0.915 \\
        \bottomrule
        \end{tabular}
    }
  \end{center}
  \label{tbl:declip}
\end{table}

\subsection{Effectiveness of NyTT in the declipping task}
We conducted experimental evaluations of NyTT in the declipping task.
Table~\ref{tbl:declip} shows the evaluation results under different $\mathrm{SNR}_{\bm{x}}$ conditions.
As in the denoising and dereverberation tasks, we can see that NyTT is effective in the declipping task, NyTT works without satisfying the Noise2Noise conditions, and the performance of NyTT improves with higher target signal quality.

\begin{figure}[t!]
  \centering
  \includegraphics[width=0.98\columnwidth]{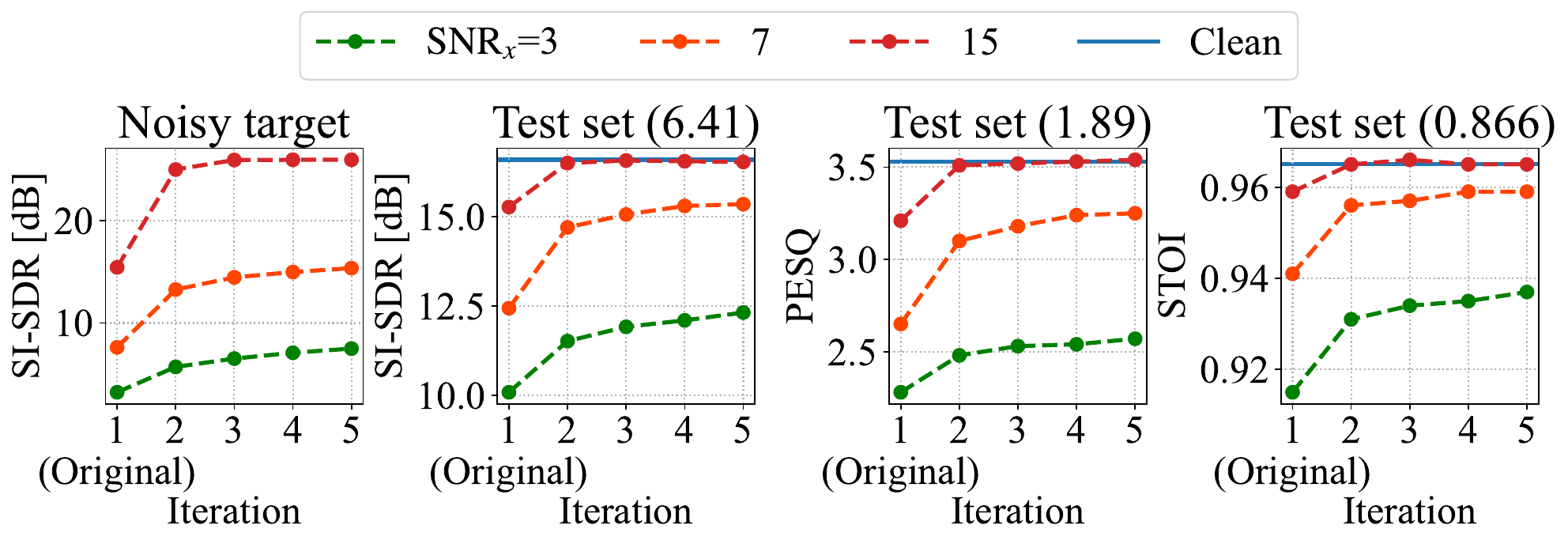}
  \caption{
    Changes in SI-SDR of the target signals and evaluation results on the test dataset through IterNyTT under different $\mathrm{SNR}_{\bm{x}}$ conditions.
  The first iteration of IterNyTT is equivalent to the original NyTT.
  Values in parentheses indicate the evaluation results of unprocessed input signals.}
\label{fig:declip_iternytt}
\end{figure}

\subsection{Effectiveness of IterNyTT in the declipping task}
To verify the effectiveness of IterNyTT in the declipping task, we evaluated the performance over five iterations under different $\mathrm{SNR}_{\bm{x}}$ conditions.
Figure~\ref{fig:declip_iternytt} illustrates the SI-SDR of the clipped targets, along with the SI-SDR, PESQ, and STOI of the processed results for the test dataset at each iteration of IterNyTT.
Here, we can again observe a consistent trend: IterNyTT improves performance, although its effectiveness is affected by the quality of the target signals.
Specifically, when $\mathrm{SNR}_{\bm{x}}$ is 15~dB, IterNyTT achieves performance comparable to that of CTT.

\section{Conclusion}
\label{sec:conclusion}
%, thereby acquiring the TSE feature.
%enhancement processing to the noisy target signals using a DNN pre-trained by NyTT, and confirmed its effectiveness.
In this study, we conducted comprehensive experimental analyses of NyTT to elucidate its detailed properties.
Our experiments revealed the following key findings:
1) NyTT can be interpreted as a training method to estimate the noisy target $\bm{x}$ by removing $\bm{n}^{\rm add}$, rather than strictly adhering to the Noise2Noise framework.
This indicates that the Noise2Noise conditions (i.e., the zero-mean distribution assumption for $\bm{n}^{\rm obs}$ and the use of the MSE loss function) are not necessary, demonstrating the flexibility of NyTT.
2) IterNyTT improved performance by enhancing the quality of noisy target signals, demonstrating its potential to achieve performance comparable to that of CTT.
3) By investigating the effects of noise mismatches, we derived desirable noise conditions.
%The desirable noise conditions were derived through the investigation of the effects of mismatches between each noise.
4) Even when a small number of clean target signals were available, the combined use of noisy and clean target signals improved performance.
% We also considered a situation where clean target signals are also available, and confirmed the effectiveness of the joint use of noisy and clean target signals.
5) NyTT was also effective in the dereverberation and declipping tasks.
Furthermore, both NyTT and IterNyTT exhibited similar behaviors across the denoising, dereverberation, and declipping tasks, implying their general applicability.

\section*{Acknowledgements}
%This work was financially supported by JST SPRING, Grant Number JPMJSP2125. The author would like to take this opportunity to thank the ``THERS Make New Standards Program for the Next Generation Researchers''.
This work was partly supported by JST CREST Grant Number JPMJCR19A3, JSPS KAKENHI Grant Number JP20H00102, and JST SPRING Grant Number JPMJSP2125.

%BACKMATTER SEE DOCUMENTATION
\section*{References}
\printbibliography

\end{document}